\documentclass[useAMS,usenatbib]{mn2e}

\usepackage{float}

\title[Solar wind dominance over the P-R effect]
  {Solar wind dominance over the Poynting-Robertson effect in secular orbital evolution of dust particles}

\author[J.~Kla\v{c}ka]
  {J.~Kla\v{c}ka \\
  Faculty of Mathematics, Physics, and Informatics, Comenius University, Mlynsk\'{a} dolina, 842 48 Bratislava, Slovak Republic \\
e-mail: klacka@fmph.uniba.sk}
\date{Released 2013 xxx XX}

\pagerange{\pageref{firstpage}--\pageref{lastpage}} \pubyear{2012}

\begin{document}

\label{firstpage}

\maketitle

\begin{abstract}
Properties of the solar wind are discussed and applied to the effect of the wind
on motion of bodies in the Solar System. The velocity density function for the solar wind
constituents is given by the $\kappa-$distribution. The relevant contributions to the solar
wind action contain also the sputtering and reflection components in addition to direct impact.
The solar wind effect is more important than the action of the solar electromagnetic
radiation, as for the secular orbital evolution.
The effect of the solar corpuscular radiation is more important than the Poynting-Robertson effect
even when mass of the dust particle is considered to be constant, non-radial component of the solar wind
velocity is neglected and the time dependence of the solar wind properties is ignored.

The presented equation of motion of a body under the action of the solar radiation,
electromagnetic and corpuscular,
respects reality in a much better way than the conventionally used equation. The acceleration
of the body is proportional to the superposition of the radial velocity component
multiplied by the numerical coefficient $[2 + (\eta_{1} + \eta_{2})/ \overline{Q} ~'_{pr}]$
and the transversal velocity component multiplied by the numerical coefficient
$(1 + \eta_{2}/ \overline{Q} ~'_{pr})$, where $\overline{Q} ~'_{pr}$ is
the dimensionless efficiency factor of the radiation pressure. Here $\eta_{1}$ $\doteq$ 1.1,
$\eta_{2}$ $\doteq$ 1.4 and the velocity is the body's velocity with respect to the Sun.
Also time variability of $\eta_{1}$ and $\eta_{2}$ due to the solar cycle is given.

The dimensionless cross section the dust grain presents to wind pressure is about 4.7.
This value differs from the conventionally used value 1.0.
The mass-loss rate of the zodiacal cloud is 4-times higher than the currently accepted value,
as for the micron-sized dust particles.
\end{abstract}


\section{Introduction}
Various non-gravitational forces have to be taken into account when
dealing with orbital evolution of dust particles in the Solar System.
Solar corpuscular radiation, i.e. solar wind, plays an important role
besides the effect of solar electromagnetic radiation. Conventional approach considers the solar
wind effect to be  (20-30)\% contribution to the velocity dependent part of the solar
electromagnetic radiation effect (e.g., Whipple 1955, 1967, Dohnanyi 1978,
Abe 2009).
This simple idea may not correspond to reality, in general (Kla\v{c}ka 1994, Bruno et al. 2003,
Kla\v{c}ka et al. 2008, 2012, Kla\v{c}ka 2013).

This paper discusses the importance of the solar wind effect on the motion of an interplanetary dust
particle (IDP). The paper may be considered as an another paper trying to understand action of various
non-gravitational effects on motion of the IDPs. While the action of the electromagnetic radiation
on spherical bodies, known as the Poynting-Robertson (P-R) effect, is well understood now
(Poynting 1903, Robertson 1937, Kla\v{c}ka 1992b, 2004, 2008a, 2008b, Kla\v{c}ka et al. 2009).
Current understanding of the P-R effect is exact and it respects both the Lorenz-Mie solution of Maxwell's equations
(Lorenz 1890, Mie 1908) and the relativity theory.
While the action of the electromagnetic radiation is completely understood now,
even for arbitrarily shaped dust grains, this paper shows that the action of the solar wind cannot be considered to be
satisfactorily understood. We reconsider the conventional statement that the
P-R effect is (3 - 5)-times more important than the solar wind effect, as for the long-term
evolution of the IDP. Although Kla\v{c}ka (2013) has shown that the P-R effect is only 1.5-times
more important than the action of the solar wind, the result holds for the Maxwell-Boltzmann velocity distribution
of the solar wind corpuscles. This paper will got further and it will consider $\kappa-$distribution as a
more realistic velocity distribution of the solar wind corpuscles.

This paper discusses the importance of the solar wind effect on motion of an IDP.
We will concentrate on the fact that the kappa-distribution is the relevant velocity distribution
for the solar wind corpuscles, in difference from the conventional approach considering the Maxwell-Boltzmann
velocity distribution. Also the fact that the relevant contributions of the solar wind action
on the IDP contain also the sputtering and reflection components in addition to direct impact, will be taken into account.

Action of an interstellar gas flow is relevant in the outer planetary zone and beyond it
(P\'{a}stor et al. 2011). We will not treat the interstellar gas flow in our paper
(we are constraining our analysis to low gas density regimes).

Stellar winds exist in greater part of stars (see, e.g., Strubbe and Chiang 2006, Plavchan et. al 2009).
Thus, better understanding of the action of the solar wind on evolution of bodies in the Solar System
can improve our understanding of dust dynamics in disks around the stars.

\section{Solar wind induced force}
We are interested in the solar wind induced force acting on an IDP.
The relative velocity of an individual solar wind particle with respect to the solar wind rest frame is
\begin{equation}\label{w}
\overrightarrow{w} = \overrightarrow{u} - \overrightarrow{u_0} ~,
\end{equation}
where the vector $\overrightarrow{u}$ is the velocity of the individual solar wind particle with respect to the Sun and the vector
$\overrightarrow{u_0} \equiv \langle\overrightarrow{u}\rangle$ is the solar wind bulk velocity or the mean/average solar wind velocity
for a given moment (we do not consider the oscillation over the solar cycle, now).
The relative velocity of the individual solar wind particle with respect to the IDP is
\begin{equation} \label{v_rel}
\overrightarrow{v_{rel}} = \overrightarrow{u} - \overrightarrow{v} = \overrightarrow{w} - (\overrightarrow{v} - \overrightarrow{u_0})  ~,
\end{equation}
where $\overrightarrow{v}$ is the velocity of the IDP with respect to the Sun.

The momentum transferred per impact is
\begin{equation} \label{hybnost}
\xi ~m_i ~\overrightarrow{v_{rel}} ~,
\end{equation}
where $\xi$ is the adsorption coefficient or sticking factor, which describes the actual type of collision
(i.e., specular, or diffuse, or adsorption; $\xi \in \langle1, 2\rangle$) and $m_i$ is the mass of a solar wind particle
(i.e., proton, electron, alpha-particle $He^{2+}$, ...).

The solar wind flux is
\begin{eqnarray}\label{tok}
\Phi_{swi} (\varphi, \overrightarrow{r}) &=& n_{i} (\varphi, \overrightarrow{r}) ~|\overrightarrow{v_{rel}}| ~,
\nonumber\\
n_{i}(\varphi, \overrightarrow{r}) &=& n_{i0}(\varphi, \hat{\overrightarrow{r}})
\left ( \frac{r_{0}}{r} \right )^{2} \equiv n_{i0} \left ( \frac{r_{0}}{r} \right )^{2} ~,
\nonumber \\
r_{0} &\equiv& 1~AU ~,
\end{eqnarray}
where $n_{i0}(\varphi, \hat{\overrightarrow{r}})$ is the local concentration of solar wind particles at 1 $AU$ and it is a function of
the solar cycle phase $\varphi$ and heliocentric (unit) position vector $\hat{\overrightarrow{r}}$, in general.
After multiplying by the cross-sectional area $\pi R^2$ of the IDP of radius $R$ we get the number of solar wind particles
\begin{equation} \label{cely tok}
\pi R^2 ~n_{i}(\varphi, \overrightarrow{r}) ~|\overrightarrow{v_{rel}}|
\end{equation}
interacting with the IDP per second.

Multiplication of expressions (\ref{hybnost}) and (\ref{cely tok})
\begin{equation} \label{cela hybnost}
\xi ~m_i ~\overrightarrow{v_{rel}} ~\pi R^2 ~n_{i}(\varphi, \overrightarrow{r}) ~|\overrightarrow{v_{rel}}|
\end{equation}
describes a change of momentum of the IDP after all interactions with solar wind particles per second.
The solar wind induced force generated by the $i-$th type of the solar wind particles leads to the acceleration
\begin{eqnarray}\label{sila vetra}
\overrightarrow{F_i} &=& \xi ~\frac{m_i}{m} ~\pi R^2 ~n_{i}(\varphi, \overrightarrow{r})
\nonumber \\
& & \times \int\limits^\infty_{-\infty} \int\limits^\infty_{-\infty} \int\limits^\infty_{-\infty} ~|\overrightarrow{v_{rel}}| ~\overrightarrow{v_{rel}} ~f_{i}(\overrightarrow{w}) ~d^3 \overrightarrow{w} ~,
\nonumber \\
\overrightarrow{v_{rel}} &=& \overrightarrow{w} - \overrightarrow{l} ~,
\nonumber \\
\overrightarrow{l} &\equiv& \overrightarrow{v} - \overrightarrow{u_0}  ~,
\end{eqnarray}
where $f_{i}(\overrightarrow{w})$ is a density function describing some velocity distribution and $m$ is mass of the body under the action of the solar
wind (interplanetary dust particle).

If the solar wind particles would behave as particles of an ideal gas, then one should use the Maxwell velocity distribution
\begin{equation}\label{Maxwell}
f_{i}^{M} (\overrightarrow{w}) = \left( \frac{m_{i}}{2 \pi k T_{M i}} \right)^{3/2} \exp \left( - \frac{m_{i} \overrightarrow{w}^2}{2 k T_{M i}} \right) ~,
\end{equation}
where $k$ is the Boltzmann constant, $m_{i}$ is mass of the ideal gas particle and $T_{M i}$ is the Maxwellian temperature. However,
the physical kinetic model of the solar wind is based on the generalized Lorentzian or $\kappa-$distribution
density function for solar wind particles (Vasyliunas 1968, Scudder 1992a, 1992b, Maksimovic \emph{et al.} 1997,
Pierrard {\it et al.} 2004, Gloeckler {\it et al.} 2010, Lazar {\it et al.} 2012):
\begin{eqnarray}\label{kappa}
f_{i}^{\kappa} (\overrightarrow{w}) &=& \frac{1}{(\pi \kappa_{i} ~w^2_{\kappa i})^{3/2}} ~
\frac{\Gamma (\kappa_{i} + 1)}{\Gamma (\kappa_{i} - 1/2)}
\nonumber \\
& & \times \left( 1 + \frac{\overrightarrow{w}^2}{\kappa_{i} ~w^2_{\kappa i}} \right)^{- (\kappa_{i} + 1)} ~,
\end{eqnarray}
where $\Gamma(x)$ is the gamma function and $w_{\kappa i}$ is an equivalent thermal speed.
The $w_{\kappa i}$ is related to the Maxwellian (thermal) temperature $T_M$ by
\begin{equation}\label{w_kappa}
w_{\kappa i}= \sqrt{\frac{(2 \kappa_{i} - 3) ~k ~T_{M i}}{\kappa_{i} ~m_{i}}} ~.
\end{equation}
We remind that $\lim\limits_{\kappa \rightarrow \infty} f^\kappa(\overrightarrow{w})$ $=$
$f^{M}(\overrightarrow{w})$.

In order to find the acceleration $\overrightarrow{F_i}$, we have to insert Eqs. (\ref{Maxwell})-(\ref{w_kappa}) into Eq. (\ref{sila vetra}).
One easily obtains (compare Banaszkiewicz {\it et al.} 1994 - Appendix B)
\begin{eqnarray} \label{F}
\overrightarrow{F_i} &=& \frac{1}{m} ~
\Lambda_i ~ (\overrightarrow{I_{1i}} - I_{2i} \overrightarrow{l} ) ~,
\end{eqnarray}
where
\begin{eqnarray}\label{lambda}
\Lambda_{i} &\equiv& \xi ~m_i^{5/2} ~\pi^{-1/2} ~R^2 ~n_{i}(\varphi, \overrightarrow{r})
\nonumber \\
& & \times \left( \frac{1}{2 k T_{Mi}} \right)^{3/2} ~,~~Maxwell~distr. ~,
\nonumber \\
\Lambda_{i} &\equiv& \xi ~m_i ~\pi^{-1/2} ~R^2 ~n_{i}(\varphi, \overrightarrow{r})
\nonumber \\
& & \times \frac{1}{(\kappa_{i} ~w^2_{\kappa i})^{3/2}} ~
\frac{\Gamma (\kappa_{i} + 1)}{\Gamma (\kappa_{i} - 1/2)} ~,~~\kappa-distr. ~,
\nonumber \\
\overrightarrow{l} &\equiv& \overrightarrow{v} - \overrightarrow{u_0}
\end{eqnarray}
and
\begin{eqnarray}\label{I1I2-0}
\overrightarrow{I_{1i}} &=& \int\limits^\infty_{-\infty} \int\limits^\infty_{-\infty} \int\limits^\infty_{-\infty}
| \overrightarrow{w} - \overrightarrow{l} | ~ \overrightarrow{w} ~g_{i} ( \overrightarrow{w} ) ~d^{3}  \overrightarrow{w}  ~,
\nonumber\\
I_{2i} &=& \int\limits^\infty_{-\infty} \int\limits^\infty_{-\infty} \int\limits^\infty_{-\infty}
|  \overrightarrow{w} - \overrightarrow{l} | ~g_{i} ( \overrightarrow{w} ) ~d^{3}  \overrightarrow{w} ~,
\nonumber\\
g_{i} (\overrightarrow{w}) &=& \exp \left ( - \frac{m_i ~\overrightarrow{w}^2}{2 k T_{Mi}} \right ) ~,~Maxwell~distr. ~,
\nonumber \\
g_{i} (\overrightarrow{w}) &=& \left( 1 + \frac{\overrightarrow{w}^2}{\kappa_{i} ~w^2_{\kappa i}} \right)^{- (\kappa_{i} + 1)} ~,~~\kappa-distr. ~.
\end{eqnarray}
Taking $\overrightarrow{l}$ $=$ $| \overrightarrow{l} |$ $\hat{\overrightarrow{l}}$ $\equiv$ $l$ $\hat{\overrightarrow{l}}$, we have
$|  \overrightarrow{w} - \overrightarrow{l} |$ $=$ $\sqrt{w^2 + l^2 - 2 ~w ~l ~\cos \vartheta'}$, where we used $w$ $=$ $| \overrightarrow{w} |$ and
$\overrightarrow{w}$ $\cdot$ $\hat{\overrightarrow{l}}$ $=$ $w ~\cos \vartheta'$. Thus, using spherical polar coordinates
in velocity space $w$ $\in$ $\langle 0, \infty )$, $\vartheta '$ $\in$ $\langle 0, \pi \rangle$, $\varphi '$ $\in$ $\langle 0, 2 \pi )$ and the
orthonormal vectors $\hat{\overrightarrow{e}}_{1}$, $\hat{\overrightarrow{e}}_{2}$, $\hat{\overrightarrow{l}}$, we have
$\overrightarrow{w}$ $=$ $w$ $\sin{\vartheta '}$ $\cos{\varphi '}$ $\hat{\overrightarrow{e}}_{1}$ $+$
$w$ $\sin{\vartheta '}$ $\sin{\varphi '}$ $\hat{\overrightarrow{e}}_{2}$ $+$ $w$ $\cos{\vartheta '}$ $\hat{\overrightarrow{l}}$
and $d^{3}  \overrightarrow{w}$ $=$ $w^2 ~\sin \vartheta' ~dw ~d\varphi' ~d\vartheta'$. Eqs. (\ref{I1I2-0}) can be rewritten,
using substitution and per partes methods, to the form
\begin{eqnarray}\label{I1I2-1}
\overrightarrow{I_{1i}} &=& 2~ \pi ~\int \limits^{\infty}_{0} \mbox{d}w ~\int\limits^{\pi}_{0} \mbox{d} \vartheta ' \left [ w^{3} ~g_{i}(w) ~ \right .
\nonumber \\
& & \left . \times ~
\sqrt{w^2 + l^2 - 2 ~w ~l ~\cos \vartheta'} ~\cos \vartheta' ~\sin \vartheta' \right ] ~\hat{\overrightarrow{l}}
\nonumber \\
&=& 2~ \pi ~\int\limits^{\infty}_{0}  w^{3} ~g_{i}(w)
\int \limits^{1}_{-1} x
\nonumber \\
& & \times ~\sqrt{w^2 + l^2 - 2 ~w ~l ~x} ~\mbox{d}x ~\mbox{d}w ~\hat{\overrightarrow{l}}
\nonumber\\
&=& -~\frac{2 \pi}{3} ~\int\limits^{\infty}_{0} w^{2} ~g_{i}(w) \left [
\left | w - l \right |^{3} + \left ( w + l \right )^{3}  \right .
\nonumber \\
& & \left . +~\frac{\left | w - l \right |^{5} - \left ( w + l \right )^{5}}{5~ w~ l} \right ] ~\mbox{d}w ~\frac{\hat{\overrightarrow{l}}}{l}
\nonumber \\
&=& -~\frac{4 \pi}{15} ~\int\limits^{l}_{0} w^{4} \left ( 5 - \frac{w^{2}}{l^{2}} \right ) ~g_{i}(w) ~\mbox{d}w ~\hat{\overrightarrow{l}}
\nonumber \\
& & -~ \frac{4 \pi}{15} ~l~\int\limits^{\infty}_{l} w^{3} \left ( 5 - \frac{l^{2}}{w^{2}} \right ) ~g_{i}(w) ~\mbox{d}w ~\hat{\overrightarrow{l}} ~,
\nonumber \\
I_{2i} &=& 2 \pi \int\limits^{\infty}_{0}  w^{2} g_{i}(w)
\int\limits^{1}_{-1} \sqrt{w^2 + l^2 - 2 ~w ~l ~x} ~\mbox{d}x \mbox{d}w
\nonumber\\
&=& -~\frac{2 \pi}{3} l^{-1} \int\limits^{\infty}_{0} w g_{i}(w) \left [
\left | w - l \right |^{3} - \left ( w + l \right )^{3} \right ] dw
\nonumber \\
&=& \frac{4 \pi}{3} ~l~\int\limits^{l}_{0} w^{2} \left ( 3 + \frac{w^{2}}{l^{2}} \right ) ~g_{i}(w) ~\mbox{d}w
\nonumber \\
& & +~ \frac{4 \pi}{3} ~\int\limits^{\infty}_{l} w^{3} \left ( 3 + \frac{l^{2}}{w^{2}} \right ) ~g_{i}(w) ~\mbox{d}w  ~,
\nonumber \\
g_{i} (\overrightarrow{w}) &=& \exp \left ( - \frac{m_i ~\overrightarrow{w}^2}{2 k T_{Mi}} \right ) ~,~~Maxwell~distr. ~,
\nonumber \\
g_{i} (\overrightarrow{w}) &=& \left( 1 + \frac{\overrightarrow{w}^2}{\kappa_{i} ~w^2_{\kappa i}} \right)^{- (\kappa_{i} + 1)} ~,~~\kappa-distr. ~.
\end{eqnarray}

\subsection{Maxwell velocity distribution}
As for the Maxwell velocity distribution, Eqs. (\ref{I1I2-1}) yield
\begin{eqnarray}\label{I1I2-M1}
\overrightarrow{I_{1i}} &=& \frac{\pi}{2} ~\left\{ - ~\frac{1}{l} ~\left( \frac{m_i}{2 k T_{Mi}} \right)^{-3} ~\exp \left( - \frac{m_i ~l^2}{2 k T_{Mi}} \right) \right.
\nonumber\\
& & \left. +~ \sqrt{\pi} ~erf \left( l ~\sqrt{\frac{m_i}{2 k T_{Mi}}} \right) \right .
\nonumber \\
& & \left . \times ~\left[ \frac{1}{2 ~l^2} ~\left( \frac{m_i}{2 k T_{Mi}} \right)^{-7/2}
- \left( \frac{m_i}{2 k T_{Mi}} \right)^{-5/2} \right] \right\} \hat{\overrightarrow{l}} ~,
\nonumber \\
I_{2i} &=& \frac{\pi}{l} ~\left( \frac{m_i}{2 k T_{Mi}} \right)^{-3/2} ~times
\nonumber \\
& & \left[ l ~\left( \frac{m_i}{2 k T_{Mi}} \right)^{-1/2}
~\exp \left( - \frac{m_i ~l^2}{2 k T_{Mi}} \right) \right.
\nonumber \\
& & \left. +~ \sqrt{\pi} ~erf \left( l ~\sqrt{\frac{m_i}{2 k T_{Mi}}} \right) ~\left( \frac{k T_{Mi}}{m_i} ~+~ l^2 \right) \right] ~,
\end{eqnarray}
where
\begin{eqnarray}\label{erf}
erf \left( x \right) &=& \frac{2}{\sqrt{\pi}} ~\int_{0}^{x} \exp \left ( -~t^{2} \right ) ~dt
\end{eqnarray}
is the error function.

Eqs. (\ref{F}), (\ref{lambda}) and (\ref{I1I2-M1}) yield for the acceleration due to the solar wind particles
\begin{eqnarray}\label{sw-force}
\overrightarrow{F} &=& -~ \frac{\pi ~ R^{2} ~\sum_{i} m_{i} ~  n_{i0} ~c_{Di}}{m~\xi}
~ \left ( \frac{r_{0}}{r} \right )^{2} ~|\overrightarrow{v} - \overrightarrow{u_0}|^{2} ~ \hat{\overrightarrow{l}}  ~,
\nonumber \\
\hat{\overrightarrow{l}} &=& (\overrightarrow{v} - \overrightarrow{u_0}) / |\overrightarrow{v} - \overrightarrow{u_0}| ~,
\nonumber \\
i &=& e, ~p, ~\alpha, ... ~,
\nonumber \\
r_{0} &=& 1 ~AU ~,
\end{eqnarray}
where $r$ is the heliocentric distance,
the symbols $p$, $\alpha$ and $e$ denote protons, He$^{2+}$ and electrons,
$c_{Di}$ is the drag coefficient given by
\begin{eqnarray}\label{c-D}
c_{Di} &=& \frac{\xi}{M_{i}} \left ( 1 + \frac{1}{2 M_{i}^{2}} \right ) \frac{e^{- M_{i}^{2}}}{\sqrt{\pi}}
\nonumber \\
& & +~ \xi \left ( 1 + \frac{1}{M_{i}^{2}} - \frac{1}{4 M_{i}^{4}} \right ) erf  \left ( M_{i} \right )
\end{eqnarray}
and
\begin{equation}\label{M}
M_{i} = |\overrightarrow{v} - \overrightarrow{u_0}| / \sqrt{\frac{2 k T_{Mi}}{m_i}}
\end{equation}
is the Mach number, $u_{0}$ $=$ 450 (468) $km~s^{-1}$ (Hundhausen 1997, Zirker 1981) and $\xi$ is the adsorption coefficient or
sticking factor which describes the actual type of collisions, $\xi$ $=$ 2 for adsorption.
Eq. (\ref{sw-force}) differs from Eqs. (19)-(20) and Eq. (A9) presented by Banaszkiewicz \emph{et al.} (1994).

Rewriting Eq. (\ref{sw-force}) into the form
\begin{eqnarray}\label{sw-force-1}
\overrightarrow{F} &=& \frac{\pi ~ R^{2} ~ m_{p} ~  n_{p0}}{m~\xi} ~ \left ( \sum_{i}
\frac{m_{i} ~  n_{i0}}{m_{p} ~  n_{p0} ~} ~c_{Di} \right )
\nonumber \\
& & \times ~ \left ( \frac{r_{0}}{r} \right )^{2}
~ | \overrightarrow{u_0} - \overrightarrow{v} | ~( \overrightarrow{u_0} - \overrightarrow{v} )  ~,
\nonumber \\
i &=& e, ~p, ~\alpha, ... ~,
\nonumber \\
r_{0} &=& 1 ~AU ~,
\end{eqnarray}
helps us to obtain the form corresponding to Banaszkiewicz \emph{et al.} (1994):
 \begin{eqnarray}\label{sw-force-2}
\overrightarrow{F} &=& \frac{\pi ~ R^{2} ~ m_{p} ~  n_{p0}}{m~\xi} ~ \left ( \sum_{i = p, \alpha}
c^{d}_{D, i} \right ) ~ \left ( \frac{r_{0}}{r} \right )^{2}
\nonumber \\
& & ~\times | \overrightarrow{u_0} - \overrightarrow{v} | ~( \overrightarrow{u_0} - \overrightarrow{v} ) ~,
\nonumber \\
r_{0} &=& 1 ~AU ~,
\end{eqnarray}
\begin{equation}\label{sw-force-pom1}
c^{d}_{D, \alpha} = \frac{m_{\alpha} ~ n_{\alpha 0}}{m_{p} ~  n_{p0} ~} ~c^{d}_{D, p}  ~.
\end{equation}
since the acceleration generated by the solar wind electrons and other ions is negligible, see
Banaszkiewicz \emph{et al.} (1994 - p. 373). The terms $c^{d}_{D, i}$ correspond to
direct impact.

If also sputtering and reflections (besides the direct impact) are considered, then
\begin{eqnarray}\label{sw-force-3}
\overrightarrow{F} &=& \frac{\pi ~ R^{2} ~ m_{p} ~  n_{p0}}{m~\xi} ~
c^{total}_{D} ~ \left ( \frac{r_{0}}{r} \right )^{2}
~ | \overrightarrow{u_0} - \overrightarrow{v} | ~( \overrightarrow{u_0} - \overrightarrow{v} ) ~,
\nonumber \\
r_{0} &=& 1 ~AU ~,
\end{eqnarray}
where
\begin{eqnarray}\label{sw-force-pom2}
c^{total}_{D} &=& \sum_{i} \left (
c^{d}_{D, i} + c^{s}_{D, i} +c^{r}_{D, i} \right )
\nonumber \\
i &=& e, ~p, ~\alpha, ...
\end{eqnarray}
and the indices $d$, $r$ and $s$ refer to direct impact, sputtering and reflection, respectively.
Conventionally, $c_{D, p}^d = 2$ for the adsorption coefficient $\xi = 2$.

As for the direct impact, we can write
\begin{eqnarray}\label{c-D-p}
c_{D, p}^d ~+~ c_{D, \alpha}^d &=&  c_{D}^d ~  ( 1 + \alpha_{d} ) ~,
\nonumber \\
\alpha_{d} &=&  \frac{n_{\alpha} ~m_{\alpha}}{n_{p} ~m_{p}} ~,
\nonumber \\
c_{D, e}^d &=& c_{D~ e}^d ~ \zeta_{d}  ~,
 \nonumber \\
\zeta_{d} &=&  \frac{n_{e} ~m_{e}}{n_{p} ~m_{p}}  ~.
\end{eqnarray}

Using the values presented in Appendix C in Banaszkiewicz {\it et al.} (1994 - p. 373), we can write
\begin{eqnarray}\label{c-D-s}
c_{D, p}^s ~+~ c_{D, \alpha}^s &=&  c_{D}^d ~  ( 0.15 + \alpha_{s} ) ~,
\nonumber \\
\alpha_{s} &\in& \langle 10, 15 \rangle ~ \frac{n_{\alpha}}{n_{p}}  ~,
\end{eqnarray}
for various materials, and,
\begin{eqnarray}\label{c-D-r}
c_{D, p}^r ~+~ c_{D, \alpha}^r &=&  c_{D}^d ~  ( 0.01 + \alpha_{r} ) ~,
\nonumber \\
\alpha_{r} &=& 0.0015  ~.
\end{eqnarray}
Moreover, it is assumed that
\begin{eqnarray}\label{c-D-e-sr}
c_{D, e}^s &=& 0 ~,
\nonumber \\
c_{D, e}^r &=& 0  ~.
\end{eqnarray}

\subsection{$\kappa-$distribution}
It is well-known that $\kappa-$distribution match the observational data on solar wind particles in much better way than the
Maxwell-Boltzmann distribution (see, e.g., Maksimovic \emph{et al.} 1997 - Fig. 3, Lazar {\it et al.} 2012).
The constituents of the solar wind are: electrons,
protons, helium nuclei ($\alpha$-particles), heavy ions and molecular ions (carbon, nitrogen, oxygen, neon, sodium,
magnesium, argon, sulphur, potassium, silicon, iron, CH$^{+}$, NH$^{+}$, OH$^{+}$, H$_{2}$O$^{+}$, H$_{3}$O$^{+}$,
MgH$^{+}$, HCN$^{+}$, C$_{2}$H$_{4}^{+}$, SO$^{+}$ and many others),
see, e.g., Pierrard {\it et al.} (2004), Gloeckler {\it et al.} (2010). The constituents are
characterized by the values $\kappa$ $\in$ $\langle 2, 5 )$ (e.g., Lazar {\it et al.} 2012).

The solar wind-induced acceleration has the following form, in the case of the $\kappa-$distribution of the solar wind particles,
see Eqs. (\ref{w_kappa}), (\ref{F}), (\ref{lambda}) and (\ref{I1I2-1}):
\begin{eqnarray}\label{F kappa-1}
\overrightarrow{F}_i &=& \frac{1}{m} ~\Lambda_{\kappa i} ~(\overrightarrow{I_{1 \kappa i}} - I_{2 \kappa i} \overrightarrow{l} ) ~,
\nonumber \\
\Lambda_{\kappa i} &\equiv& \xi ~m_i ~\pi^{-1/2} ~R^2 ~n_{i}(\varphi, \overrightarrow{r})
\nonumber \\
& & \times \frac{1}{(\kappa_i ~w^2_{\kappa i})^{3/2}} ~\frac{\Gamma (\kappa_i + 1)}{\Gamma (\kappa_i - 1/2)} ~,
\nonumber \\
\overrightarrow{l} &\equiv& \overrightarrow{v} - \overrightarrow{u_0} ~,
\nonumber \\
\overrightarrow{I_{1 \kappa i}} &=& -~ \frac{4 \pi}{15} ~\int\limits^l_0 ~
w^{4} \left ( 5 -  \frac{w^2}{l^2} \right )
\nonumber \\
& & \times \left( 1 + \frac{w^2}{\kappa_i ~w^2_{\kappa i}} \right)^{- (\kappa_i + 1)} ~\mbox{d}w ~\hat{\overrightarrow{l}}
\nonumber\\
&& -~ \frac{4 \pi}{15} ~l ~\int\limits^\infty_l ~w^{3} \left ( 5 -  \frac{l^2}{w^2} \right )
\nonumber \\
& & \times \left( 1 + \frac{w^2}{\kappa_i ~w^2_{\kappa i}} \right)^{- (\kappa_i + 1)} ~\mbox{d}w ~\hat{\overrightarrow{l}} ~,
\nonumber \\
I_{2 \kappa i} &=& \frac{4 \pi}{3} ~l \int\limits^l_0 w^2 \left ( 3 + \frac{w^2}{l^2} \right )
\left( 1 + \frac{w^2}{\kappa_i ~w^2_{\kappa i}} \right)^{- (\kappa_i + 1)} \mbox{d}w
\nonumber \\
&& + \frac{4 \pi}{3} \int\limits^\infty_l w^3 \left ( 3 + \frac{l^2}{w^2} \right )
\left( 1 + \frac{w^2}{\kappa_i ~w^2_{\kappa i}} \right)^{- (\kappa_i + 1)} \mbox{d}w ~,
\nonumber \\
w_{\kappa i} &=& \sqrt{(2 \kappa_{i} - 3) k T_{M i} / ( \kappa_{i} m_{i} )} ~,
\nonumber \\
\hat{\overrightarrow{l}} &=& \overrightarrow{l} / | \overrightarrow{l} | ~.
\end{eqnarray}
The acceleration represented by Eqs. (\ref{F kappa-1}) corresponds to the direct impact component. Thus, the total
acceleration can be written as
\begin{eqnarray}\label{sw-force-total-hi2}
\overrightarrow{F} &\doteq& \frac{\pi R^{2} m_{p} n_{p0}}{m~\xi} \left ( \frac{r_{0}}{r} \right )^{2}
| \overrightarrow{u_0} - \overrightarrow{v} | ( \overrightarrow{u_0} - \overrightarrow{v} ) ~c_D^{tot} ~,
\nonumber \\
c_D^{tot} &=& \sum_{j} \left ( c_{D, j}^d ~+~ c_{D, j}^s ~+~ c_{D, j}^r \right ) ~,
\nonumber \\
c_{D, j}^d &=& \frac{4~ \xi ^{2}}{\sqrt{\pi}}  ~\frac{\Gamma (\kappa_{j} + 1)}{\Gamma (\kappa_{j} - 1/2)} ~ \frac{c_{D, j}^d (M)}{c_{D}^d} ~
\times X_{j}^d ~,
\nonumber \\
c_{D, j}^d (M) &=& c_{D}^d ~ \frac{n_{j} ~m_{j}}{n_{p} ~m_{p}} ~,
\nonumber \\
c_{D, p}^s &=& 0.15 ~c_{D}^d  ~,
\nonumber \\
c_{D, e}^s &\doteq& 0 ~,
\nonumber \\
c_{D, j}^s &=& \langle 2.50, 3.75 \rangle \times c_{D}^d ~ \frac{n_{j} ~m_{j}}{n_{p} ~m_{p}} ~, ~~ if ~j \ne ~ \mbox{e, p} ~,
\nonumber \\
c_{D, j}^r &=& 0.01 ~c_{D}^d ~ \frac{n_{j} ~m_{j}}{n_{p} ~m_{p}} ~,
\nonumber \\
\xi &=& 2 ~,
\nonumber \\
c_{D}^d &=& 2 ~,
\nonumber \\
r_{0} &=& 1 ~\mbox{AU} ~,
\nonumber \\
j &=& \mbox{e, p}, ~ ^{4}\mbox{He}, ~\mbox{heavy ions, molecular ions} ~,
\end{eqnarray}
where also Eqs. (\ref{sw-force-pom1}), (\ref{c-D-p}), (\ref{c-D-s}), (\ref{c-D-r}) and (\ref{c-D-e-sr}) are used
(experimental data may improve the numerical values, in future), and,
\begin{eqnarray}\label{sw-force-total-hi2b}
X_{j}^d &=& \frac{4}{3} ~ q_{j}^{3} ~ \int_{1}^{\infty} \frac{x / 5 + x^{3}}{\left ( 1 + q_{j}^{2} ~x^{2} \right )^{\kappa_{j} + 1}} ~\mbox{d}x
\nonumber \\
& & +~ q_{j}^{3} \int_{0}^{1} \frac{x^{2} + 2 x^{4} / 3 -  x^{6} / 15}{\left ( 1 + q_{j}^{2} ~x^{2} \right )^{\kappa_{j} + 1}} ~\mbox{d}x ~,
\nonumber \\
q_{j} &=& | \overrightarrow{u_0} - \overrightarrow{v} | / ( \sqrt{\kappa_j} ~w_{\kappa j} ) ~,
\nonumber \\
w_{\kappa j} &=& \sqrt{(2 \kappa_{j} - 3) ~k ~T_{M, j} / ( \kappa_{j} ~m_{j} ) } ~.
\end{eqnarray}

The first integral in Eqs. (\ref{sw-force-total-hi2b}) can be easily calculated using substitution
and per partes methods:
\begin{eqnarray}\label{sw-hi01}
I_{\infty, j} &\equiv& \int_{1}^{\infty} \frac{x / 5 + x^{3}}{\left ( 1 + q_{j}^{2} ~x^{2} \right )^{\kappa_{j} + 1}} ~\mbox{d}x
\nonumber \\
&=& \frac{3}{5~\kappa_{j}} ~\frac{1}{q_{j}^{2} \left ( 1 + q_{j}^{2} \right )^{\kappa_{j}}}
\left [ 1 + \frac{5 \left ( 1 + 1 / q_{j}^{2} \right )}{6~\left ( \kappa_{j} - 1 \right)}  \right ] ~.
\end{eqnarray}

Similarly, the second integral in Eqs. (\ref{sw-force-total-hi2b}) yields
\begin{eqnarray}\label{sw-hi02}
I_{0, j} &\equiv& \int_{0}^{1} \frac{x^{2} + 2 x^{4} / 3 - x^{6} / 15}{\left ( 1 + q_{j}^{2} ~x^{2} \right )^{\kappa_{j} + 1}} ~ \mbox{d}x
\nonumber \\
&=& \frac{1}{2 \kappa_{j}} \left [ \frac{1}{q_{j}^{3}} ~ J \left ( \kappa_{j}, q_{j} \right ) ~-~ \frac{8}{5} ~
\frac{1}{q_{j}^{2} \left ( 1 + q_{j}^{2} \right ) ^{\kappa_{j}}} \right ]
\nonumber \\
& & + ~\frac{1}{2 \kappa_{j} \left ( \kappa_{j} - 1 \right )} ~ \frac{1}{q_{j}^{5}} ~
J \left ( \kappa_{j} - 1, q_{j} \right )
\nonumber \\
& & -~ \frac{5}{12 \kappa_{j} \left ( \kappa_{j} - 1 \right )} \frac{1}{q_{j}^{4} \left ( 1 + q_{j}^{2} \right ) ^{\kappa_{j} - 1}}
\nonumber \\
& & -~ \frac{1}{4 \kappa_{j} \left ( \kappa_{j} - 1 \right )}
\frac{1}{q_{j}^{7}}
\nonumber \\
& & \times \left [ J \left ( \kappa_{j} - 2, q_{j} \right ) - J \left ( \kappa_{j} - 1, q_{j} \right ) \right ]  ~,
\nonumber \\
J \left ( \zeta, q \right ) &\equiv& \int_{0}^{q} ~\left ( 1 + z^{2} \right )^{-~\zeta} ~\mbox{d}z ~.
\end{eqnarray}

Eqs. (\ref{sw-force-total-hi2b}), (\ref{sw-hi01}) and (\ref{sw-hi02}) yield
\begin{eqnarray}\label{sw-hi03}
X_{j}^d &=& \frac{1}{4} ~ \frac{1}{\kappa_{j} \left ( \kappa_{j} - 1 \right)} ~\frac{1}{q_{j} \left ( 1 + q_{j}^{2} \right )^{\kappa_{j} - 1}} +
\nonumber \\
& & \frac{1}{2}~ \frac{1}{\kappa_{j}} ~ J \left ( \kappa_{j}, q_{j} \right ) +
\nonumber \\
& & \frac{1}{2} \frac{1}{\kappa_{j} \left ( \kappa_{j} - 1 \right )}  \frac{1}{q_{j}^{2}}
\left ( 1 + \frac{1}{2}  \frac{1}{q_{j}^{2}} \right ) J \left ( \kappa_{j} - 1, q_{j} \right )
\nonumber \\
& & -~ \frac{1}{4} ~\frac{1}{\kappa_{j} \left ( \kappa_{j} - 1 \right )} ~ \frac{1}{q_{j}^{4}} ~J \left ( \kappa_{j} - 2, q_{j} \right ) ~,
\nonumber \\
J \left ( \zeta, q \right ) &\equiv& \int_{0}^{q} ~\left ( 1 + z^{2} \right )^{-~\zeta} ~\mbox{d}z ~,
\nonumber \\
J \left ( \zeta, q \right ) &=&  \frac{2 \zeta - 3}{2 \left ( \zeta - 1 \right )} ~J \left ( \zeta - 1, q \right )
\nonumber \\
& & +~ \frac{1}{2 \left ( \zeta - 1 \right )} ~ \frac{q}{\left ( 1 + q^{2} \right ) ^{\zeta - 1}} ~,
\nonumber \\
q_{j} &=& | \overrightarrow{u_0} - \overrightarrow{v} | / ( \sqrt{\kappa_j} ~w_{\kappa j} ) ~,
\nonumber \\
w_{\kappa j} &=& \sqrt{(2 \kappa_{j} - 3) ~k ~T_{M, j} / ( \kappa_{j} ~m_{j} ) } ~.
\end{eqnarray}

\section{Observational data}
Average values for protons, $\alpha$-particles and electrons are collected in Table 1 for two different sources.

\begin{table}
\begin{center}
\begin{tabular}{|c|c|c|}
\hline
$quantity$ & Zirker & Hundhausen    \\
\hline
$n_{p 0}$ [ $cm^{-3}$ ]      & 8.70  & 6.60  \\
$n_{\alpha 0}$ [ $cm^{-3}$ ] & 0.34  & 0.25  \\
$n_{e 0}$ [ $cm^{-3}$ ]      & 9.38  & 7.10  \\
$u_{0}$ [ $km~s^{-1}$ ]      & 468   & 450  \\
\hline
\end{tabular}
\caption{Characteristics of the solar wind, according to the data presented by Zirker (1981 -- Tables 5-3 and 5-4)
and Hundhausen (1997 -- p. 92). Average values of the concentrations of protons, He$^{2+}$ and electrons (at 1 AU),
together with the average value of the wind speed are presented.}
\end{center}
\label{tab:1}
\end{table}

Tables 2 and 3 present relative concentrations of several solar wind constituents for two sets of
observational data. Table 4 offers temperature (measured in SI units) of several solar wind ions.
Since we do not have temperatures of $^{20}$Ne and $^{24}$Mg
in disposal, we consider two different values in this paper:
$T[^{20}$Ne$^{1)}]$ $=$ 6.905 $\times$ 10$^{-16}$ $J$,
$T[^{24}$Mg$^{1)}]$ $=$ 6.905 $\times$ 10$^{-16}$ $J$, and,
$T[^{20}$Ne$^{2)}]$ $=$ 2.762 $\times$ 10$^{-15}$ $J$,
$T[^{24}$Mg$^{2)}]$ $=$ 2.762 $\times$ 10$^{-15}$ $J$.
In any case, low concentrations of $^{20}$Ne and $^{24}$Mg secure that the temperatures of the
elements do not play any significant role in motion of a body under the action of the solar wind.

\begin{table}
\begin{center}
\begin{tabular}{|c|c|}
\hline
$j$ & $n_{j}/n_{p}$   \\
\hline
\hline
e                & 1.076                                    \\
p                & 1.000                                    \\
$^{4}\mbox{He}$  & 3.788 $\times$ 10$^{-2}$                 \\
$^{16}\mbox{O}$  & (5.354 $\pm$ 0.916) $\times$ 10$^{-4}$   \\
$^{20}\mbox{Ne}$ & (6.693 $\pm$ 1.029) $\times$ 10$^{-5}$   \\
$^{24}\mbox{Mg}$ & (8.031 $\pm$ 1.742) $\times$ 10$^{-5}$   \\
\hline
\end{tabular}
\caption{Relative concentrations of several solar wind constituents. The input data are taken from Lazar {\it et al.} (2012),
$http://ulysses.jpl.nasa.gov/science/mission\_primary.html$ and Hundhausen (1997), see also Table 1.}
\end{center}
\label{tab:2}
\end{table}

\begin{table}
\begin{center}
\begin{tabular}{|c|c|}
\hline
$j$ & $n_{j}/n_{p}$  \\
    &                \\
\hline
\hline
e                & 1.078                                    \\
p                & 1.000                                    \\
$^{4}\mbox{He}$  & 3.908 $\times$ 10$^{-2}$                 \\
$^{16}\mbox{O}$  & (4.062 $\pm$ 0.695) $\times$ 10$^{-4}$   \\
$^{20}\mbox{Ne}$ & (5.077 $\pm$ 0.781) $\times$ 10$^{-5}$   \\
$^{24}\mbox{Mg}$ & (6.092 $\pm$ 1.322) $\times$ 10$^{-5}$   \\
\hline
\end{tabular}
\caption{Relative concentrations of several solar wind constituents.
The input data are taken from Lazar {\it et al.} (2012),
$http://ulysses.jpl.nasa.gov/science/mission\_primary.html$ and Zirker (1981), see also Table 1.}
\end{center}
\label{tab:3}
\end{table}

\begin{table}
\begin{center}
\begin{tabular}{|c|c|}
\hline
$j$ &  $k~T_{M, j}$ [$J$] \\
\hline
\hline
e                 & 1.381 $\times$ 10$^{-17}$  \\
p                 & 1.381 $\times$ 10$^{-17}$  \\
$^{4}\mbox{He}$   & 6.905 $\times$ 10$^{-16}$  \\
$^{16}\mbox{O}$   & 2.762 $\times$ 10$^{-15}$  \\
\hline
\end{tabular}
\caption{Temperatures of several solar wind ions (Pierrard 2012),
$k$ is the Boltzmann constant.}
\end{center}
\label{tab:4}
\end{table}

\section{Numerical calculations}
Eqs. (\ref{sw-force-total-hi2}) and (\ref{sw-hi03}) represent relevant equation of motion of a body due to the action of the solar wind.
We will discuss the special case, in order to show the importance of the equation of motion. We will use the value $\kappa$
$=$ 2.0, as a very good approximation to reality (see Maksimovic et al. 1997, Lazar {\it et al.} 2012, Pierrard 2012).

If $\kappa_{j}$ $=$ 2, then Eqs. (\ref{sw-force-total-hi2}) yield
\begin{eqnarray}\label{sw-force-total-hi2d}
c_{D, j}^d (\kappa = 2) &=& 32.000 \times \pi^{-1} \times X_{j}^d \times c_{D, j}^d (M) ~,
\nonumber \\
c_{D, j}^d (M) &=& 2 ~\frac{n_{j} ~m_{j}}{n_{p} ~m_{p}} ~,
\nonumber \\
j &=& \mbox{e, p}, ^{4}\mbox{He}, \mbox{heavy ions, molecular ions} ~.
\end{eqnarray}

Eqs. (\ref{sw-hi03}) reduce to
\begin{eqnarray}\label{sw-hi03-kappa2}
X_{j}^d &=& \frac{1}{8} ~\frac{1}{q_{j}^{3}} \left [ q_{j}^{-1} ~arctg \left ( q_{j} \right ) - 1 \right ]
\nonumber \\
& & + ~\frac{1}{8} ~ \frac{1}{q_{j}} ~ \left [ 2 ~q_{j}^{-1} ~arctg \left ( q_{j} \right ) + 1 \right ]
\nonumber \\
& & +~ \frac{1}{8} ~ arctg \left ( q_{j} \right ) ~,
\nonumber \\
q_{j} &=& | \overrightarrow{u_0} - \overrightarrow{v} | / ( \sqrt{2} ~w_{\kappa j} ) ~,
\nonumber \\
w_{\kappa j} &=& \sqrt{k ~T_{M, j} / ( 2 ~m_{j} ) } ~.
\end{eqnarray}

As an example, we can mention that the electron drag coefficient for the direct impact is 30-times
greater than it is conventionally assumed, i. e., $c_{D, e}^d (\kappa = 2)$ $=$ 29.4 $c_{D, e}^d (Maxwell-distribution)$.
The approximation $q_{j}$ $\doteq$ $q_{0j}$ $\equiv$ $| \overrightarrow{u_0} | / ( \sqrt{\kappa_j} ~w_{\kappa j} )$ is used
and the data of Hundhausen (1997) are taken into account.

\section{Equation of motion of IDP under the action of the solar wind}
On the basis of our presentation we can conclude that the equation of motion of a spherical body,
under the action of the solar gravity, P-R effect and the solar wind,
can be obtained from Eqs. (\ref{sw-force-total-hi2}) and (\ref{sw-force-total-hi2b}), or, using $\kappa$ $=$ 2 as a very good
approximation, Eqs. (\ref{sw-force-total-hi2d})-(\ref{sw-hi03-kappa2}). The acceleration of the body
due to the solar wind is
\begin{eqnarray}\label{sw-force-total-fin-kappa2}
\overrightarrow{F} &\doteq& \frac{\pi R^{2} m_{p} n_{p}}{2~m} ~ \left ( \frac{r_{0}}{r} \right )^{2}  ~
| \overrightarrow{u} - \overrightarrow{v} | ( \overrightarrow{u} - \overrightarrow{v} ) ~c_D^{tot} ~,
\nonumber \\
c_D^{tot} &=& \sum_{j} \left [ \alpha_{sw} \left ( 1 - \delta_{jp} \right ) +  0.3 ~\delta_{jp}
+ \frac{8}{\pi} ~\beta_{j} \right ] ~ \frac{n_{j}}{n_{p}} ~ \frac{m_{j}}{m_{p}} ~,
\nonumber \\
\beta_{j} &=& \left ( 1 + \frac{2}{q_{j}^{2}} + \frac{1}{q_{j}^{4}} \right ) arctg \left ( q_{j} \right )
+ \frac{1}{q_{j}} - \frac{1}{q_{j}^{3}}  ~,
\nonumber \\
q_{j} &=& | \overrightarrow{u} - \overrightarrow{v} | / \zeta (j)  ~,
\nonumber \\
\zeta (j) &=& \sqrt{k ~T_{M, j} / m_{j}}  ~,
\nonumber \\
\overrightarrow{u} &=& u \hat{\overrightarrow{u}} ~,
\nonumber \\
u &=& u_{0} ( 1 - \delta \cos{\varphi} ) ~,
\nonumber \\
n_{p} &=& n_{p0} ( 1 - \delta \cos{\varphi} ) ~,
\nonumber \\
\delta &=& 0.15 ~,
\nonumber \\
\varphi &=& 2 \pi ~\frac{t - t_{r} - t_{max}}{T} ~, ~~t_{r} \doteq \frac{r}{2 u_{0}} ~,~~ T = 11.1 ~yr ~,
\nonumber \\
\hat{\overrightarrow{u}} &=& \gamma_{R} ~\overrightarrow{e}_{R} ~+~ \gamma_{T} ~\hat{\overrightarrow{u}}_{T} ~,
\nonumber \\
\gamma_{R} &=& \cos{\varepsilon} ~,
\nonumber \\
\gamma_{T} &=& \sin{\varepsilon} ~,
\nonumber \\
\varepsilon &\in& \langle 2^{\circ}, 3^{\circ} \rangle ~,
\nonumber \\
\overrightarrow{e}_{R} &=& \overrightarrow{r} / | \overrightarrow{r} | \equiv \overrightarrow{r} / r ~,
\nonumber \\
\hat{\overrightarrow{u}}_{T} &=& \hat{\overrightarrow{\omega}} \times \overrightarrow{e}_{R} / | \hat{\overrightarrow{\omega}} \times \overrightarrow{e}_{R} |  ~,
\nonumber \\
\hat{\overrightarrow{\omega}} &=& ( \sin{\Omega_{S}} ~\sin{i_{S}}, -~\cos{\Omega_{S}} ~\sin{i_{S}}, \cos{i_{S}} ) ~,
\nonumber \\
i_{S} &=& 7^{\circ} 15' ~, ~~ \Omega_{S} = 73^{\circ} 40' + 50.25'' ( t [\mbox{yr}] - 1850 ) ~,
\nonumber \\
n_{p0} &\equiv& n_{p0} ( r_{0} ) ~,
\nonumber \\
r_{0} &=& 1 ~\mbox{AU} ~,
\nonumber \\
\alpha_{sw} &=& 6.2 \pm 1.2 ~,
\nonumber \\
j &=& \mbox{e, p}, ~ ^{4}\mbox{He}, ~\mbox{heavy ions, molecular ions} ~,
\end{eqnarray}
where $\delta_{ij}$ is the Kronecker delta, Tables 1 - 4 (Sec. 3) can be used, $\alpha_{sw}$ depends on material
properties of dust particle, $t_{max}$ is the instant of the solar cycle maximum, results from Kla\v{c}ka (1994)
are used for $\hat{\overrightarrow{\omega}}$ in ecliptic coordinates and ($k~T_{M, j}$ $=$ 6.905 $\times$ 10$^{-16}$ $J$ $^{1)}$ and
$k~T_{M, j}$ $=$ 2.762 $\times$ 10$^{-15}$ $J$ $^{2)}$)
\begin{eqnarray}\label{sw-force-total-fin-kappa2-zeta}
\zeta (p) &=& 9.0855 \times 10^{1} ~\mbox{km}~\mbox{s}^{-1} ~,
\nonumber \\
\zeta (e) &=& 3.8930 \times 10^{3} ~\mbox{km}~\mbox{s}^{-1} ~,
\nonumber \\
\zeta (^{4}\mbox{He}) &=& 3.2122 \times 10^{2} ~\mbox{km}~\mbox{s}^{-1} ~,
\nonumber \\
\zeta (^{16}\mbox{O}) &=& 3.2122 \times 10^{2} ~\mbox{km}~\mbox{s}^{-1} ~,
\nonumber \\
\zeta (^{20}\mbox{Ne}^{1)}) &=& 1.43654 \times 10^{2} ~\mbox{km}~\mbox{s}^{-1} ~,
\nonumber \\
\zeta (^{24}\mbox{Mg}^{1)}) &=& 1.3114 \times 10^{2} ~\mbox{km}~\mbox{s}^{-1} ~,
\nonumber \\
\zeta (^{20}\mbox{Ne}^{2)}) &=& 2.8731 \times 10^{2} ~\mbox{km}~\mbox{s}^{-1} ~,
\nonumber \\
\zeta (^{24}\mbox{Mg}^{2)}) &=& 2.6228 \times 10^{2} ~\mbox{km}~\mbox{s}^{-1} ~.
\end{eqnarray}
In general, if one wants to make more exact calculations on the basis of the future observational data,
then Secs. 2 - 4 have to be used. However, the approximation $\kappa$ $=$ 2 is a much better approximation
to reality than the conventional approach based on the Maxwell-Boltzmann velocity distribution.

\subsection{Analytical approach}
The aim of this section is to make some analytical calculations.
The derived results can immediately shed light on the simple and most important
terms in the equation of motion of the IDP under the action of the solar wind.
The new results are easily comparable with the standard results used in scientific
literature and textbooks during the last century.

\subsubsection{Some useful relations}
At first, the following expansions can be easily derived:
\begin{eqnarray}\label{expansions}
q_{j} &\equiv& | \overrightarrow{u} - \overrightarrow{v} | / \zeta (j)  ~,
\nonumber \\
q_{j} &\doteq& q_{0j} \times
\nonumber \\
& & \left [ 1 - \frac{\overrightarrow{v} \cdot \hat{\overrightarrow{u}}}{u}
+ \frac{1}{2} \left ( \frac{v}{u} \right )^{2} - \frac{1}{2} \left ( \frac{\overrightarrow{v} \cdot \hat{\overrightarrow{u}}}{u} \right )^{2}
\right ]  ~,
\nonumber \\
q_{j}^{-1} &\doteq& q_{0j}^{-1} \times
\nonumber \\
& & \left [ 1 + \frac{\overrightarrow{v} \cdot \hat{\overrightarrow{u}}}{u}
- \frac{1}{2} \left ( \frac{v}{u} \right )^{2} + \frac{3}{2} \left ( \frac{\overrightarrow{v} \cdot \hat{\overrightarrow{u}}}{u} \right )^{2}
\right ]  ~,
\nonumber \\
q_{j}^{-2} &\doteq& q_{0j}^{-2} \times
\nonumber \\
& & \left [ 1 + 2 \frac{\overrightarrow{v} \cdot \hat{\overrightarrow{u}}}{u}
- \left ( \frac{v}{u} \right )^{2} + 4 \left ( \frac{\overrightarrow{v} \cdot \hat{\overrightarrow{u}}}{u} \right )^{2}
\right ]  ~,
\nonumber \\
q_{j}^{-3} &\doteq& q_{0j}^{-3} \times
\nonumber \\
& & \left [ 1 + 3 \frac{\overrightarrow{v} \cdot \hat{\overrightarrow{u}}}{u}
- \frac{3}{2} \left ( \frac{v}{u} \right )^{2} + \frac{15}{2} \left ( \frac{\overrightarrow{v} \cdot \hat{\overrightarrow{u}}}{u} \right )^{2}
\right ]  ~,
\nonumber \\
q_{j}^{-4} &\doteq& q_{0j}^{-4} \times
\nonumber \\
& & \left [ 1 + 4 \frac{\overrightarrow{v} \cdot \hat{\overrightarrow{u}}}{u}
- 2 \left ( \frac{v}{u} \right )^{2} + 12 \left ( \frac{\overrightarrow{v} \cdot \hat{\overrightarrow{u}}}{u} \right )^{2}
\right ]  ~,
\nonumber \\
arctg \left ( q_{j} \right ) &\doteq& arctg \left ( q_{0j} \right )
 + \frac{q_{0j}}{1 + q_{0j}^{2}} \times
 \nonumber \\
& & \left [ - \frac{\overrightarrow{v} \cdot \hat{\overrightarrow{u}}}{u}
+ \frac{1}{2} \left ( \frac{v}{u} \right )^{2} - \frac{1}{2}
\left ( \frac{\overrightarrow{v} \cdot \hat{\overrightarrow{u}}}{u} \right )^{2} \right ]
\nonumber \\
& & - \frac{q_{0j}^{3}}{\left ( 1 + q_{0j}^{2} \right )^{2}} \left ( \frac{\overrightarrow{v} \cdot \hat{\overrightarrow{u}}}{u} \right )^{2} ~,
\nonumber \\
q_{0j} &\equiv& u ~\left [ \zeta (j) \right ]^{-1} ~.
\end{eqnarray}
Eqs. (\ref{expansions}) enable to find
\begin{eqnarray}\label{exp-beta}
\beta_{j} &=& \left ( 1 + \frac{2}{q_{j}^{2}} + \frac{1}{q_{j}^{4}} \right ) arctg \left ( q_{j} \right )
+ \frac{1}{q_{j}} - \frac{1}{q_{j}^{3}}
\nonumber \\
&\doteq& \left ( 1 + \frac{2}{q_{0j}^{2}} + \frac{1}{q_{0j}^{4}} \right ) arctg \left ( q_{0j} \right )
+  \frac{1}{q_{0j}}  - \frac{1}{q_{0j}^{3}}
\nonumber \\
& & + \left \{ \left ( 1 + \frac{1}{q_{0j}^{2}} \right ) \frac{4}{q_{0j}^{2}} ~arctg \left ( q_{0j} \right )
+ \frac{1}{q_{0j}} - \frac{3}{q_{0j}^{3}} \right \}
\nonumber \\
& & \times \left [ \frac{\overrightarrow{v} \cdot \hat{\overrightarrow{u}}}{u} - \frac{1}{2} ~\left ( \frac{\overrightarrow{v}}{u} \right )^{2} \right ]
\nonumber \\
& & - \left ( 1 + \frac{2}{q_{0j}^{2}} + \frac{1}{q_{0j}^{4}} \right )
\frac{q_{0j}}{1 + q_{0j}^{2}} ~
\left [ \frac{\overrightarrow{v} \cdot \hat{\overrightarrow{u}}}{u} - \frac{1}{2} ~\left ( \frac{\overrightarrow{v}}{u} \right )^{2} \right ]
\nonumber \\
& & + \left \{ \left ( 2 + \frac{3}{q_{0j}^{2}} \right ) \frac{4}{q_{0j}^{2}} ~arctg \left ( q_{0j} \right )
+ \frac{3}{2} \left ( 1 -  \frac{5}{q_{0j}^{2}} \right ) \frac{1}{q_{0j}}  \right.
\nonumber \\
& & \left. - \frac{4}{q_{0j}^{3}}
\right \}  \left ( \frac{\overrightarrow{v} \cdot \hat{\overrightarrow{u}}}{u} \right )^{2}
- \frac{q_{0j}}{1 + q_{0j}^{2}} \left ( \frac{1}{2} + \frac{q_{0j}^{2}}{1 + q_{0j}^{2}} \right )
\nonumber \\
& & \times
\left ( 1 + \frac{2}{q_{0j}^{2}} + \frac{1}{q_{0j}^{4}} \right )
\left ( \frac{\overrightarrow{v} \cdot \hat{\overrightarrow{u}}}{u} \right )^{2} ~.
\end{eqnarray}
Similarly,
\begin{eqnarray}\label{exp-u-v}
| \overrightarrow{u} - \overrightarrow{v} | ( \overrightarrow{u} - \overrightarrow{v} ) &\doteq& u^{2}
\left [ 1 - \frac{\overrightarrow{v} \cdot \hat{\overrightarrow{u}}}{u} + \frac{1}{2} ~\left ( \frac{\overrightarrow{v}}{u} \right )^{2}
\right ] \hat{\overrightarrow{u}}
\nonumber \\
& & - u^{2} ~\frac{1}{2} \left ( \frac{\overrightarrow{v} \cdot \hat{\overrightarrow{u}}}{u} \right )^{2} \hat{\overrightarrow{u}}
\nonumber \\
& & - u^{2} \left ( 1 - \frac{\overrightarrow{v} \cdot \hat{\overrightarrow{u}}}{u} \right ) \frac{\overrightarrow{v}}{u} ~.
\end{eqnarray}

\subsubsection{Acceleration - dominant terms}
Insertion of Eqs. (\ref{exp-beta})-(\ref{exp-u-v}) into Eqs. (\ref{sw-force-total-fin-kappa2}) yields
\begin{eqnarray}\label{sw-force-total-fin-kappa2-exp-tilde}
\overrightarrow{F} &\doteq& \frac{\pi R^{2} m_{p} n_{p} (r_{0})}{2~m} ~ \left ( \frac{r_{0}}{r} \right )^{2}  ~ u^{2}
\times \overrightarrow{X}_{Fsw}
\nonumber \\
\overrightarrow{X}_{Fsw} &=& \left ( \tilde{\eta}_{2} - \tilde{\eta}_{1}
\frac{\overrightarrow{v} \cdot \hat{\overrightarrow{u}}}{u} \right ) \hat{\overrightarrow{u}}
- \tilde{\eta}_{2} ~\frac{\overrightarrow{v}}{u}
\nonumber \\
& & +~ \frac{1}{2} ~\tilde{\eta}_{1}  \left ( \frac{\overrightarrow{v}}{u} \right )^{2} \hat{\overrightarrow{u}}
+ \tilde{\eta}_{1} \frac{\overrightarrow{v} \cdot \hat{\overrightarrow{u}}}{u} \frac{\overrightarrow{v}}{u}
\nonumber \\
& & -~ \frac{1}{2} ~\tilde{\eta}_{3}  \left ( \frac{\overrightarrow{v} \cdot \hat{\overrightarrow{u}}}{u} \right )^{2} \hat{\overrightarrow{u}} ~,
\end{eqnarray}
where we have omitted the terms of higher orders in $v/u$ and
\begin{eqnarray}\label{tilde-eta-1-2-3}
\tilde{\eta}_{1} &=& -  \frac{8}{\pi} \sum_{j} \left [ \left ( 1 + \frac{1}{q_{0j}^{2}} \right ) \frac{4}{q_{0j}^{2}} ~arctg \left ( q_{0j} \right )
+ \frac{1}{q_{0j}} - \frac{3}{q_{0j}^{3}} \right ]
\nonumber \\
& & \times \frac{n_{j}}{n_{p}} ~ \frac{m_{j}}{m_{p}}
\nonumber \\
& & + \frac{8}{\pi} \sum_{j} \left ( 1 + \frac{2}{q_{0j}^{2}} + \frac{1}{q_{0j}^{4}} \right )
\frac{q_{0j}}{1 + q_{0j}^{2}} ~\frac{n_{j}}{n_{p}} ~ \frac{m_{j}}{m_{p}} ~+~ \tilde{\eta}_{2} ~,
\nonumber \\
\tilde{\eta}_{2} &=& \sum_{j} \left [ \alpha_{sw} \left ( 1 - \delta_{jp} \right ) +  0.3 ~\delta_{jp} \right ] ~ \frac{n_{j}}{n_{p}} ~ \frac{m_{j}}{m_{p}}
+ \frac{8}{\pi} \times
\nonumber \\
& & \sum_{j} \left [ \left ( 1 + \frac{2}{q_{0j}^{2}} + \frac{1}{q_{0j}^{4}} \right ) arctg \left ( q_{0j} \right )
+ \frac{1}{q_{0j}} - \frac{1}{q_{0j}^{3}} \right ]
\nonumber \\
& & \times \frac{n_{j}}{n_{p}} ~ \frac{m_{j}}{m_{p}} ~,
\nonumber \\
\tilde{\eta}_{3} &=& - \frac{64}{\pi} \sum_{j} \left ( 2 + \frac{3}{q_{0j}^{2}} \right ) \frac{1}{q_{0j}^{2}} ~arctg \left ( q_{0j} \right )
\frac{n_{j}}{n_{p}} ~ \frac{m_{j}}{m_{p}}
\nonumber \\
& & + \frac{16}{\pi} \sum_{j} \left ( 1 + \frac{2}{q_{0j}^{2}} + \frac{1}{q_{0j}^{4}} \right )
\frac{q_{0j}^{3}}{\left ( 1 + q_{0j}^{2} \right )^{2}} ~\frac{n_{j}}{n_{p}} ~ \frac{m_{j}}{m_{p}}
\nonumber \\
& & + \frac{16}{\pi} \sum_{j} \frac{12 + 11 q_{0j}^{2} - q_{0j}^{4}}{q_{0j}^{3} \left ( 1 + q_{0j}^{2} \right )} \frac{n_{j}}{n_{p}} \frac{m_{j}}{m_{p}}
+3 \tilde{\eta}_{2} - 2 \tilde{\eta}_{1} ~,
\nonumber \\
q_{0j} &\equiv& u ~\left [ \zeta (j) \right ]^{-1} ~.
\end{eqnarray}
One can use the values in Tables 1 - 4, see also Eqs. (\ref{sw-force-total-fin-kappa2-zeta}). Finally,
(\ref{sw-force-total-fin-kappa2-exp-tilde})-(\ref{tilde-eta-1-2-3}) lead to the acceleration of the IDP due to the solar wind
\begin{eqnarray}\label{sw-accel-total-fin-kappa2-exp}
\overrightarrow{a}_{sw} &\doteq& \frac{\beta}{\overline{Q} ~'_{pr}} ~ \frac{G~M_{\odot}}{r^{2}} \times \overrightarrow{X}_{asw}
\nonumber \\
\overrightarrow{X}_{asw} &=& \left ( \eta_{2} ~ \frac{u}{c} ~-~
\eta_{1} ~\frac{\overrightarrow{v} \cdot \hat{\overrightarrow{u}}}{c} \right ) \hat{\overrightarrow{u}}
~-~ \eta_{2} ~\frac{\overrightarrow{v}}{c}
\nonumber \\
& & +~ \frac{1}{2} ~\eta_{1} ~\frac{\overrightarrow{v} \cdot \overrightarrow{v}}{u~c} ~ \hat{\overrightarrow{u}}
+ \eta_{1} ~\frac{\overrightarrow{v} \cdot \hat{\overrightarrow{u}}}{u} \frac{\overrightarrow{v}}{c}
\nonumber \\
& & -~ \frac{1}{2} ~\eta_{3}  ~\frac{ \left ( \overrightarrow{v} \cdot \hat{\overrightarrow{u}} \right )^{2}}{u~c} \hat{\overrightarrow{u}} ~,
\end{eqnarray}
where we introduced the dimensionless quantities (see also Kla\v{c}ka 2008a, 2008b, Kla\v{c}ka et al. 2009)
\begin{eqnarray}\label{beta-eta-kappa}
\beta &=& \frac{L_{\odot} ~R^{2} ~\overline{Q} ~'_{pr}}{4 G M_{\odot} m c} ~,
\nonumber \\
\eta_{j} &=& \frac{c^{2} m_{p} u n_{p} (r_{0})}{\xi ~S_{0}} \tilde{\eta}_{j} \equiv
\frac{c^{2} ~m_{p} ~u~ n_{p}(r_{0})}{\xi \left [ L_{\odot} / \left ( 4 \pi ~r_{0}^{2} \right ) \right ]}
~\tilde{\eta}_{j}  ~, ~ j = 1, 2, 3 ~,
\nonumber \\
\xi &=& 2 ~.
\end{eqnarray}
$L_{\odot}$ is the rate of energy outflow from the Sun, the solar luminosity, $R$ is the radius of the IDP, $m$ its mass,
$\overline{Q} ~'_{pr}$ is the dimensionless efficiency factor of the radiation pressure averaged over the solar spectrum,
$G$ is the gravitational constant, $M_{\odot}$ is the mass of the Sun, $c$ is the speed of light in vacuum
$m_{p}$ is the proton mass, $n_{p}$ is the proton concentration in the solar wind and
$S_{0}$ $=$ 1.366 $\times$ $10^{3}$ $W~m^{-2}$ denotes the solar electromagnetic flux.
Numerically,
\begin{eqnarray}\label{beta}
\beta &=& 5.760 \times 10^{2} \frac{\overline{Q} ~'_{pr}}{R (\mu m) \rho (kg~m^{-3})} ~,
\end{eqnarray}
where $R$ and $\rho$ are radius and mass density of the homogeneous spherical IDP, and,
\begin{eqnarray}\label{eta-1-2-Hun-Zir}
\eta_{1} &\doteq& 1.0  ~, ~~~~\mbox{Hundhausen data} ~,
\nonumber \\
\eta_{1} &\doteq& 1.3 ~, ~~~~\mbox{Zirker data} ~,
\nonumber \\
\eta_{2} &\doteq& 1.2 ~, ~~~~\mbox{Hundhausen data} ~,
\nonumber \\
\eta_{2} &\doteq& 1.6 ~, ~~~~\mbox{Zirker data} ~,
\nonumber \\
\eta_{3} &\doteq& 0.9 ~, ~~~~\mbox{Hundhausen data} ~,
\nonumber \\
\eta_{3} &\doteq& 1.2 ~, ~~~~\mbox{Zirker data} ~.
\end{eqnarray}
The error is about 0.1.
In reality, the values presented in Eqs. (\ref{eta-1-2-Hun-Zir}) are mean values.
Time variability may be considered in the equation of motion.

On the basis of Eqs. (\ref{eta-1-2-Hun-Zir}) we can conclude that $\eta_{2}$ $>$ $\eta_{1}$ $>$ $\eta_{3}$, and, approximately,
\begin{eqnarray}\label{eta-1-2}
\eta_{10} &\doteq& 1.1  ~,
\nonumber \\
\eta_{20} &\doteq& 1.4  ~,
\nonumber \\
\eta_{30} &\doteq& 1.0  ~,
\end{eqnarray}
where the index $0$ denotes the time mean value. We used $\eta_{j0}$ $\doteq$
[$\eta_{j} (Hundhausen)$ $+$ $\eta_{j} (Zirker)$] / 2, $j$ $=$ 1, 2, 3.

It can be verified that the values presented in Eq. (\ref{eta-1-2}) hold not only for
$\kappa$ $=$ 2, but also for $\kappa$ $=$ 4. We can summarize that the results
given by Eq. (\ref{eta-1-2}) hold for 2 $\le$ $\kappa$ $<$ 5. In any case, the
observed properties of the solar wind are characterized by the values collected in Eq. (\ref{eta-1-2}).

If we take into account solar cycle, then
\begin{eqnarray}\label{ety-sol-cycle1}
\eta_1 &=& \eta_{10} \left ( 1 - \eta_{1A} ~\delta ~\cos{\varphi} \right )  ~,
\nonumber \\
\eta_2 &=& \eta_{20} \left ( 1 - \eta_{2A} ~\delta ~\cos{\varphi} \right )  ~,
\nonumber \\
\eta_3 &=& \eta_{30} \left ( 1 - \eta_{3A} ~\delta ~\cos{\varphi} \right )  ~,
\end{eqnarray}
where Eqs. (\ref{sw-force-total-fin-kappa2})-(\ref{sw-force-total-fin-kappa2-zeta}) are used,
the mean values are given by Eqs. (\ref{eta-1-2}) and the amplitudes, describing an importance
of the oscillations due to the existence of the solar cycle, are
\begin{eqnarray}\label{ety-sol-cycle2}
\eta_{1A} &\doteq& 1.9 ~,
\nonumber \\
\eta_{2A} &\doteq& 2.2 ~,
\nonumber \\
\eta_{3A} &\doteq& 1.8 ~.
\end{eqnarray}

If we want to take into account the change of the solar wind properties due to the existence of the solar cycle, then we may be
interested in the contribution of the variable component in comparison with the mean value of the accelerations.
The amplitude of the dominant part of the variable solar wind acceleration is proportional to
$( \eta_{20} / \overline{Q} ~'_{pr} ) \eta_{2A} \delta$, since the transversal component
is more relevant than the radial component. The mean acceleration is
proportional to $( 1 + \eta_{20} / \overline{Q} ~'_{pr} )$. The ratio
$( \eta_{20} / \overline{Q} ~'_{pr} ) \eta_{2A} \delta / ( 1 + \eta_{20} / \overline{Q} ~'_{pr} )$
is presented in Table 5, for several combinations of the values of the physical parameters
$\eta_{20}$, $\eta_{2A}$  and $\overline{Q} ~'_{pr}$
(time variation was considered in Kla\v{c}ka et al. 2012, so the corresponding values
$\eta_{20}$ $=$ 0.38 and $\eta_{2A}$ $=$ 2.0 are taken as reference values; $\delta$ $=$ 0.15).
The greater the value of $\eta_{20}$, the greater is the importance of the variable solar
wind component, for a given value of $\overline{Q} ~'_{pr}$. In general,
($\eta_{20} / \overline{Q} ~'_{pr}) \eta_{2A} \delta / ( 1 + \eta_{20} / \overline{Q} ~'_{pr} )$ $<$ $\eta_{2A} \delta$ $=$ 0.33,
since $\eta_{2A}$ $=$ 2.2 and $\delta$ $=$ 0.15.

\begin{table}
\begin{center}
\begin{tabular}{|c|c| c c c|}
\hline
$\eta_{20}$ & $\eta_{2A}$ & & $\eta_{20} \eta_{2A} \delta / ( \eta_{20} + \overline{Q} ~'_{pr} )$  &   \\
\hline
         &  & $\overline{Q} ~'_{pr} = 1/2$  & $\overline{Q} ~'_{pr} = 1$  & $\overline{Q} ~'_{pr} = 2$  \\
\hline
\hline
0.38  & 2.0 & 0.130  & 0.083  & 0.048  \\
1.40  & 2.2 & 0.243  & 0.193  & 0.136  \\
\hline
\end{tabular}
\caption{The ratio of the amplitude of the variable solar wind acceleration to the mean value of the acceleration, i.e.,
$\eta_{20} \eta_{2A} \delta / ( \eta_{20} + \overline{Q} ~'_{pr} )$,
for several values of the constants $\eta_{20}$ and $\overline{Q} ~'_{pr}$.}
\end{center}
\label{tab:5}
\end{table}

If one would take into account that $n_{i}$ $=$ $n_{i 0}$ $( 1 - \delta \cos \varphi )$, $i$ $=$ $p$,
$\alpha$, $e$, and, $\delta$ $\in$ $\langle$ 0.14, 0.15 $\rangle$, $\varphi$ changes in 2 $\pi$ within a solar cycle
(Kla\v{c}ka et al. 2012), then the ratios between the Zirker and the Hundhausen data correspond to, approximately,
$(1 + \delta)$ / $(1 - \delta)$. This would indicate that the Zirker data hold for the solar cycle minimum
and the Hundhausen data hold for the solar cycle maximum. The  Zirker and the Hundhausen data show that
although the $n_{i}$ $=$ $n_{i 0}$ $( 1 - \delta \cos{\varphi} )$ holds for solar wind particles,
the speed of the particles is practically constant, independent on the solar cycle: $u$ $=$ $u_{0}$ $\in$
$\langle 450, 468 \rangle$ $\mbox{km} ~\mbox{s}^{-1}$. One may use the value 450 $\mbox{km} ~\mbox{s}^{-1}$,
or, the mean value 460 $\mbox{km} ~\mbox{s}^{-1}$. However, other observational data suggest
$u$ $=$ $u_{0}$ $( 1 - \delta \cos \varphi )$ (e.g., Banaszkiewicz et al. 1994).

\subsubsection{Solar radiation and equation of motion of IDP}
Equation of motion of the spherical IDP under the action of the solar electromagnetic and corpuscular
radiation is
\begin{eqnarray}\label{accel-total-1st-order1}
\frac{d \overrightarrow{v}}{dt} &\doteq& \beta ~\frac{G~M_{\odot}}{r^{2}}
\left ( \overrightarrow{e}_{R}
+ \frac{\eta_{2}}{\overline{Q} ~'_{pr}} ~ \frac{u}{c} ~\hat{\overrightarrow{u}} \right )
\nonumber \\
& & -~ \beta ~\frac{G~M_{\odot}}{r^{2}} \left (
\frac{\overrightarrow{v} \cdot \overrightarrow{e}_{R}}{c} ~ \overrightarrow{e}_{R}
~+~ \frac{\eta_{1}}{\overline{Q} ~'_{pr}} ~
\frac{\overrightarrow{v} \cdot \hat{\overrightarrow{u}}}{c}  ~\hat{\overrightarrow{u}} \right )
\nonumber \\
& & -~ \beta ~\frac{G~M_{\odot}}{r^{2}} \left (  1 ~+~ \frac{\eta_{2}}{\overline{Q} ~'_{pr}} \right ) \frac{\overrightarrow{v}}{c}
\nonumber \\
& & +  \beta \frac{G M_{\odot}}{r^{2}} \left [ \frac{1}{2} \frac{\eta_{1}}{\overline{Q}'_{pr}}
\frac{\overrightarrow{v} \cdot \overrightarrow{v}}{u~c} \hat{\overrightarrow{u}}
+ \frac{\eta_{1}}{\overline{Q}'_{pr}} ~\frac{\overrightarrow{v} \cdot \hat{\overrightarrow{u}}}{u} \frac{\overrightarrow{v}}{c}
\right ]
\nonumber \\
& & -~  \beta ~\frac{G~M_{\odot}}{r^{2}}  \frac{1}{2} ~\frac{\eta_{3}}{\overline{Q} ~'_{pr}}
~\frac{ \left ( \overrightarrow{v} \cdot \hat{\overrightarrow{u}} \right )^{2}}{u~c} ~\hat{\overrightarrow{u}}  ~,
\end{eqnarray}
where Eq. (\ref{sw-accel-total-fin-kappa2-exp}) is used, and, see Eqs. (\ref{sw-force-total-fin-kappa2}),
\begin{eqnarray}\label{auxiliary1}
\overrightarrow{u} &=& u \hat{\overrightarrow{u}} ~,
\nonumber \\
u &=& u_{0} ( 1 - \delta \cos{\varphi} )~,
\nonumber \\
\delta &=& 0.15 ~,
\nonumber \\
\varphi &=& 2 \pi ~\frac{t - t_{r} - t_{max}}{T} ~, ~~t_{r} \doteq \frac{r}{2 u_{0}} ~,~~ T = 11.1 ~yr ~,
\nonumber \\
\hat{\overrightarrow{u}} &=& \gamma_{R} ~\overrightarrow{e}_{R} ~+~ \gamma_{T} ~\hat{\overrightarrow{u}}_{T} ~,
\nonumber \\
\gamma_{R} &=& \cos{\varepsilon} ~,
\nonumber \\
\gamma_{T} &=& \sin{\varepsilon} ~,
\nonumber \\
\varepsilon &\in& \langle 2^{\circ}, 3^{\circ} \rangle ~,
\nonumber \\
\overrightarrow{e}_{R} &=& \overrightarrow{r} / | \overrightarrow{r} | \equiv \overrightarrow{r} / r ~,
\nonumber \\
\hat{\overrightarrow{u}}_{T} &=& \hat{\overrightarrow{\omega}} \times \overrightarrow{e}_{R} / | \hat{\overrightarrow{\omega}} \times \overrightarrow{e}_{R} |  ~,
\nonumber \\
\hat{\overrightarrow{\omega}} &=& ( \sin{\Omega_{S}} ~\sin{i_{S}}, -~\cos{\Omega_{S}} ~\sin{i_{S}}, \cos{i_{S}} ) ~,
\nonumber \\
u_{0} &=& 450 ~km~s^{-1} ~,
\nonumber \\
i_{S} &=& 7^{\circ} 15' ~, ~~ \Omega_{S} = 73^{\circ} 40' + 50.25'' ( t [yr] - 1850 )
\end{eqnarray}
in ecliptic coordinates, and, see Eqs. (\ref{beta}), (\ref{eta-1-2}), (\ref{ety-sol-cycle1}), (\ref{ety-sol-cycle2}),
\begin{eqnarray}\label{auxiliary2}
\beta &=& 5.760 \times 10^{2} \frac{\overline{Q} ~'_{pr}}{R (\mu m) \rho (kg~m^{-3})} ~,
\nonumber \\
\eta_1 &=& \eta_{10} \left ( 1 - \eta_{1A} ~\delta ~\cos{\varphi} \right )  ~,
\nonumber \\
\eta_2 &=& \eta_{20} \left ( 1 - \eta_{2A} ~\delta ~\cos{\varphi} \right )  ~,
\nonumber \\
\eta_3 &=& \eta_{30} \left ( 1 - \eta_{3A} ~\delta ~\cos{\varphi} \right )  ~,
\nonumber \\
\eta_{10} &\doteq& 1.1  ~,
\nonumber \\
\eta_{20} &\doteq& 1.4  ~,
\nonumber \\
\eta_{30} &\doteq& 1.0  ~,
\nonumber \\
\eta_{1A} &\doteq& 1.9 ~,
\nonumber \\
\eta_{2A} &\doteq& 2.2 ~,
\nonumber \\
\eta_{3A} &\doteq& 1.8 ~.
\end{eqnarray}

Using the approximation $\hat{\overrightarrow{u}}$ $\doteq$ $\overrightarrow{e}_{R}$, Eq. (\ref{accel-total-1st-order1}) reduces to
the following simple form:
\begin{eqnarray}\label{accel-total-1st-order2}
\frac{d \overrightarrow{v}}{dt} &\doteq& \beta ~\frac{G~M_{\odot}}{r^{2}} \times
\nonumber \\
& & \left [ 1 + \frac{\eta_{2}}{\overline{Q} ~'_{pr}} ~ \frac{u}{c}
~-~ \left ( 1 +  \frac{\eta_{1}}{\overline{Q} ~'_{pr}} \right )
\frac{\overrightarrow{v} \cdot \overrightarrow{e}_{R}}{c}  \right ] \overrightarrow{e}_{R}
\nonumber \\
& & +~ \beta ~\frac{G~M_{\odot}}{r^{2}} \frac{1}{2} \times
\nonumber \\
& & \left [ \frac{\eta_{1}}{\overline{Q} ~'_{pr}}
~\frac{\overrightarrow{v} \cdot \overrightarrow{v}}{u~c} - \frac{\eta_{3}}{\overline{Q} ~'_{pr}}
~\frac{ \left ( \overrightarrow{v} \cdot \hat{\overrightarrow{e}}_{R} \right )^{2}}{u~c} \right ]  \hat{\overrightarrow{e}}_{R}
\nonumber \\
& & -~ \beta ~\frac{G~M_{\odot}}{r^{2}} \times
\nonumber \\
& & \left ( 1 ~+~ \frac{\eta_{2}}{\overline{Q} ~'_{pr}}
- \frac{\eta_{1}}{\overline{Q} ~'_{pr}} ~\frac{\overrightarrow{v} \cdot \hat{\overrightarrow{e}}_{R}}{u}
\right ) \frac{\overrightarrow{v}}{c} ~.
\end{eqnarray}
The term $( \eta_{2} / \overline{Q} ~'_{pr} ) u/ c$ cannot be neglected, in general. It can be neglected for the constant solar wind,
since its value is small in comparison with 1. However, for the (time-)variable solar wind, the variable term can be dominant
with respect to other variable terms caused by the solar wind (see also Kla\v{c}ka et al. 2012).

\subsubsection{Discussion}
If one would like to use the fact that the solar wind speed depends on the phase of the solar cycle (Banaszkiewicz et al. 1994, Kla\v{c}ka et al. 2012),
then
\begin{eqnarray}\label{auxiliary1}
u &=& u_{0} ( 1 - \delta \cos \varphi ) ~,
\nonumber \\
\varphi &=& 2 \pi ~\frac{t - t_{r} - t_{max}}{T} ~,
\nonumber \\
t_{r} &\doteq& \frac{r}{2 u_{0}} ~,~~ T = 11.1 ~yr ~,
\nonumber \\
u_{0} &=& (450-470) ~km~s^{-1} ~,
\end{eqnarray}
can be used. We remind that the quantity $t_{max}$ is the instant of the solar cycle maximum.
Eqs. (\ref{auxiliary2}) hold in this case.

If the solar wind speed would not depend on the phase of the solar cycle, then
\begin{eqnarray}\label{auxiliary2-n}
u &=& u_{0}  ~,
\nonumber \\
\varphi &=& 2 \pi ~\frac{t - t_{r} - t_{max}}{T} ~,
\nonumber \\
t_{r} &=& \frac{r}{u_{0}} ~,~~ T = 11.1 ~yr ~,
\nonumber \\
u_{0} &=& (450-470) ~km~s^{-1} ~.
\end{eqnarray}
should be used. Moreover, Eqs. (\ref{auxiliary2}) have to be replaced by the following equations:
\begin{eqnarray}\label{auxiliary2-speed-non-solar-cycle}
\beta &=& 5.760 \times 10^{2} \frac{\overline{Q} ~'_{pr}}{R (\mu m) \rho (kg~m^{-3})} ~,
\nonumber \\
\eta_1 &=& \eta_{10} \left ( 1 - \eta_{1A} ~\delta ~\cos{\varphi} \right )  ~,
\nonumber \\
\eta_2 &=& \eta_{20} \left ( 1 - \eta_{2A} ~\delta ~\cos{\varphi} \right )  ~,
\nonumber \\
\eta_3 &=& \eta_{30} \left ( 1 - \eta_{3A} ~\delta ~\cos{\varphi} \right )  ~,
\nonumber \\
\eta_{10} &\doteq& 1.1  ~,
\nonumber \\
\eta_{20} &\doteq& 1.4  ~,
\nonumber \\
\eta_{30} &\doteq& 1.0  ~,
\nonumber \\
\eta_{1A} &=& 1.0 ~,
\nonumber \\
\eta_{2A} &=& 1.0 ~,
\nonumber \\
\eta_{3A} &=& 1.0 ~.
\end{eqnarray}

\subsubsection{Secular orbital evolution}
In order to better understand the action of the solar wind, we will neglect time variability of the wind, i.e. we will put
$\delta$ $=$ 0 in Eqs. (\ref{sw-force-total-fin-kappa2}). This assumption enables us to make some simple analytical calculations. We will use Eqs. (\ref{accel-total-1st-order1})
with the approximation
\begin{eqnarray}\label{sec-orb-1}
\hat{\overrightarrow{u}} &=& \overrightarrow{e}_{R} ~+~ \gamma_{T} ~\overrightarrow{e}_{T} ~,
\nonumber \\
\gamma_{T} &=&
\frac{\hat{\overrightarrow{\omega}} \times \overrightarrow{e}_{R}}{| \hat{\overrightarrow{\omega}} \times \overrightarrow{e}_{R} |}
\cdot \overrightarrow{e}_{T} ~ \sin{\varepsilon}  ~,
\nonumber \\
\hat{\overrightarrow{\omega}} &=& ( 0, 0, 1) ~,
\nonumber \\
\overrightarrow{e}_{T} &=& \frac{
\overrightarrow{v} - \left ( \overrightarrow{v} \cdot \overrightarrow{e}_{R} \right ) \overrightarrow{e}_{R}}{
| \overrightarrow{v} - \left ( \overrightarrow{v} \cdot \overrightarrow{e}_{R} \right ) \overrightarrow{e}_{R} |} ~,
\nonumber \\
\varepsilon &\in& \langle 2^{\circ}, 3^{\circ} \rangle ~.
\end{eqnarray}
The vector $\hat{\overrightarrow{\omega}}$ can be used in ecliptic coordinates, approximately, see Eqs. (\ref{sw-force-total-fin-kappa2}).

Let us consider motion of a spherical IDP in the gravitational field of the Sun and under the action of the solar radiation.
On the basis of Eqs. (\ref{accel-total-1st-order1}) and (\ref{sec-orb-1}) we can write
\begin{eqnarray}\label{accel-total-1st-order3}
\frac{d \overrightarrow{v}}{dt} &\doteq& -~\frac{G~M_{\odot}}{r^{2}} ~\overrightarrow{e}_{R}
\nonumber \\
& & +~ \beta ~\frac{G~M_{\odot}}{r^{2}}
\left [ 1 ~-~ \left ( 1 ~+~ \frac{\eta_{1}}{\overline{Q} ~'_{pr}} \right )
\frac{\overrightarrow{v} \cdot \overrightarrow{e}_{R}}{c} \right ] \overrightarrow{e}_{R}
\nonumber \\
& & -~ \beta ~\frac{G~M_{\odot}}{r^{2}} ~\gamma_{T} ~\frac{\eta_{1}}{\overline{Q} ~'_{pr}}
\frac{\overrightarrow{v} \cdot \overrightarrow{e}_{T}}{c} ~ \overrightarrow{e}_{R}
\nonumber \\
& & +~ \beta ~\frac{G~M_{\odot}}{r^{2}} ~\gamma_{T}
\left ( \frac{\eta_{2}}{\overline{Q} ~'_{pr}} \frac{u}{c} - \frac{\eta_{1}}{\overline{Q} ~'_{pr}}
\frac{\overrightarrow{v} \cdot \overrightarrow{e}_{R}}{c} \right ) \overrightarrow{e}_{T}
\nonumber \\
& & -~ \beta ~\frac{G~M_{\odot}}{r^{2}} \left ( 1 ~+~ \frac{\eta_{2}}{\overline{Q} ~'_{pr}} \right ) \frac{\overrightarrow{v}}{c}
\nonumber \\
& & +~  \beta ~\frac{G~M_{\odot}}{r^{2}}  \frac{1}{2} ~\frac{\eta_{1}}{\overline{Q} ~'_{pr}}
~\frac{\overrightarrow{v} \cdot \overrightarrow{v}}{u~c} ~ \overrightarrow{~e}_{R}
\nonumber \\
& & -~  \beta ~\frac{G~M_{\odot}}{r^{2}}  \frac{1}{2} ~\frac{\eta_{3}}{\overline{Q} ~'_{pr}}
~ \frac{\left ( \overrightarrow{v} \cdot \overrightarrow{e}_{R} \right )^{2}}{u~c}
~\overrightarrow{e}_{R}
\nonumber \\
& & -~  \beta ~\frac{G~M_{\odot}}{r^{2}} ~\frac{\eta_{3}}{\overline{Q} ~'_{pr}}
\gamma_{T}
\frac{\left ( \overrightarrow{v} \cdot \overrightarrow{e}_{R} \right ) \left ( \overrightarrow{v} \cdot \overrightarrow{e}_{T} \right )}{u~c} ~\overrightarrow{e}_{R}
\nonumber \\
& & +~ \beta ~\frac{G~M_{\odot}}{r^{2}} ~\gamma_{T} ~\frac{1}{2} ~\frac{\eta_{1}}{\overline{Q} ~'_{pr}}
~\frac{\overrightarrow{v} \cdot \overrightarrow{v}}{u~c} ~ \overrightarrow{~e}_{T}
\nonumber \\
& & -~ \beta ~\frac{G~M_{\odot}}{r^{2}} ~\gamma_{T} ~\frac{1}{2} ~ \frac{\eta_{3}}{\overline{Q} ~'_{pr}}
~\frac{ \left ( \overrightarrow{v} \cdot \overrightarrow{e}_{R} \right )^{2}}{u~c} ~ \overrightarrow{~e}_{T}
\nonumber \\
& & +~ \beta ~ \frac{G~M_{\odot}}{r^{2}}  \frac{\eta_{1}}{\overline{Q} ~'_{pr}} ~
\nonumber \\
& & \times \left ( \frac{\overrightarrow{v} \cdot \overrightarrow{e}_{R}}{u} + \gamma_{T} \frac{\overrightarrow{v} \cdot \overrightarrow{e}_{T}}{u}
\right ) \frac{\overrightarrow{v}}{c} ~,
\end{eqnarray}
if the terms containing $\gamma_{T}^{2}$ are neglected, or,
\begin{eqnarray}\label{accel-total-1st-order4}
\frac{d \overrightarrow{v}}{dt} &\doteq& -~\frac{G~M_{\odot} \left ( 1 - \beta \right )}{r^{2}} ~\overrightarrow{e}_{R} ~-~ \beta ~\frac{G~M_{\odot}}{r^{2}}
\times
\nonumber \\
& & \left [ \left ( 1 + \frac{\eta_{1}}{\overline{Q} ~'_{pr}} \right )
\frac{\overrightarrow{v} \cdot \overrightarrow{e}_{R}}{c}
+ \gamma_{T} \frac{\eta_{1}}{\overline{Q}'_{pr}}
\frac{\overrightarrow{v} \cdot \overrightarrow{e}_{T}}{c} \right ] \overrightarrow{e}_{R}
\nonumber \\
& & +~ \beta ~\frac{G~M_{\odot}}{r^{2}} ~\gamma_{T}
\left ( \frac{\eta_{2}}{\overline{Q} ~'_{pr}} \frac{u}{c} - \frac{\eta_{1}}{\overline{Q} ~'_{pr}}
\frac{\overrightarrow{v} \cdot \overrightarrow{e}_{R}}{c} \right ) \overrightarrow{e}_{T}
\nonumber \\
& & -~ \beta ~\frac{G~M_{\odot}}{r^{2}} \left ( 1 ~+~ \frac{\eta_{2}}{\overline{Q} ~'_{pr}} \right ) \frac{\overrightarrow{v}}{c}
\nonumber \\
& & +~  \beta ~\frac{G~M_{\odot}}{r^{2}}  \frac{1}{2} ~\frac{\eta_{1}}{\overline{Q} ~'_{pr}}
~\frac{\overrightarrow{v} \cdot \overrightarrow{v}}{u~c} ~ \overrightarrow{~e}_{R}
\nonumber \\
& & -~  \beta ~\frac{G~M_{\odot}}{r^{2}} \frac{1}{2} ~\frac{\eta_{3}}{\overline{Q} ~'_{pr}} \times
\nonumber \\
& & \left [ \frac{\left ( \overrightarrow{v} \cdot \overrightarrow{e}_{R} \right )^{2}}{u~c} + 2 \gamma_{T}
\frac{\left ( \overrightarrow{v} \cdot \overrightarrow{e}_{R} \right ) \left ( \overrightarrow{v} \cdot \overrightarrow{e}_{T} \right )}{u~c}
\right ] \overrightarrow{e}_{R}
\nonumber \\
& & +~ \beta ~\frac{G~M_{\odot}}{r^{2}} ~\gamma_{T} ~\frac{1}{2}
\nonumber \\
& & \times \left [ \frac{\eta_{1}}{\overline{Q} ~'_{pr}}
~\frac{\overrightarrow{v} \cdot \overrightarrow{v}}{u~c} ~-~ \frac{\eta_{3}}{\overline{Q} ~'_{pr}}
~\frac{ \left ( \overrightarrow{v} \cdot \overrightarrow{e}_{R} \right )^{2}}{u~c} \right ]~ \overrightarrow{~e}_{T}
\nonumber \\
& & +~ \beta ~\frac{G~M_{\odot}}{r^{2}}  \frac{\eta_{1}}{\overline{Q} ~'_{pr}} \times
\nonumber \\
& & \left ( \frac{\overrightarrow{v} \cdot \overrightarrow{e}_{R}}{u} + \gamma_{T} ~\frac{\overrightarrow{v} \cdot \overrightarrow{e}_{T}}{u}
\right ) \frac{\overrightarrow{v}}{c} ~.
\end{eqnarray}

We will not deal with secular evolution of the IDP if the central acceleration is given by the gravity of the Sun. This
can be done on the basis of Eq. (\ref{accel-total-1st-order3}) and Sec. 6.2 in Kla\v{c}ka (2004). Our approach
will concentrate on the case when the central acceleration is given both by the gravity of the Sun and the dominant part
of the radiation pressure, as it is given by the first part on the right-hand side of Eq. (\ref{accel-total-1st-order4}).
The orbital elements referring to the central acceleration $-~[G~M_{\odot} \left ( 1 - \beta \right )/r^{2}] \overrightarrow{e}_{R}$
will be characterized by the subscript $\beta$, to be consistent with Kla\v{c}ka (2004).

Eq. (\ref{accel-total-1st-order4}) yields the following radial and transversal components (normal component equals to 0)
of the perturbation acceleration to Keplerian motion:
\begin{eqnarray}\label{celmech1}
F_{\beta R} &=& - \beta ~\frac{G~M_{\odot}}{r^{2}} \times
\nonumber \\
& & \left [ \left ( 2 + \frac{\eta_{1}}{\overline{Q} ~'_{pr}} + \frac{\eta_{2}}{\overline{Q} ~'_{pr}} \right )
\frac{v_{R}}{c} + \gamma_{T} ~\frac{\eta_{1}}{\overline{Q} ~'_{pr}}
\frac{v_{T}}{c} \right ]
\nonumber \\
& & +~ \frac{1}{2} ~\beta ~\frac{G~M_{\odot}}{r^{2}} \times
\nonumber \\
& & \left [ \frac{\eta_{1}}{\overline{Q}'_{pr}} \frac{v_{R}^{2} + v_{T}^{2}}{u~c} -
\frac{\eta_{3}}{\overline{Q}'_{pr}} \left ( \frac{v_{R}^{2}}{u~c} + 2 \gamma_{T}
\frac{v_{R} v_{T}}{u~c} \right ) \right ]
\nonumber \\
& & +~  \frac{\eta_{1}}{\overline{Q} ~'_{pr}} ~
\left ( \frac{v_{R}}{u} + \gamma_{T} ~\frac{v_{T}}{u} \right ) \frac{v_{R}}{c} ~\beta ~\frac{G~M_{\odot}}{r^{2}}  ~,
\nonumber \\
F_{\beta T} &=& -~ \beta ~\frac{G~M_{\odot}}{r^{2}} \times
\nonumber \\
& & \left [ \left ( 1 + \frac{\eta_{2}}{\overline{Q}'_{pr}} \right )
\frac{v_{T}}{c} - \gamma_{T}
\left ( \frac{\eta_{2}}{\overline{Q}'_{pr}} \frac{u}{c} - \frac{\eta_{1}}{\overline{Q}'_{pr}}
\frac{v_{R}}{c} \right ) \right ]
\nonumber \\
& & +~\gamma_{T} \frac{1}{2} \left ( \frac{\eta_{1}}{\overline{Q}'_{pr}}
\frac{v_{R}^{2} + v_{T}^{2}}{u~c} - \frac{\eta_{3}}{\overline{Q}'_{pr}}
\frac{v_{R}^{2}}{u~c} \right )  \beta ~\frac{G~M_{\odot}}{r^{2}}
\nonumber \\
& & +~  \frac{\eta_{1}}{\overline{Q} ~'_{pr}} ~
\left ( \frac{v_{R}}{u} + \gamma_{T} ~\frac{v_{T}}{u} \right ) \frac{v_{T}}{c} ~\beta ~\frac{G~M_{\odot}}{r^{2}}  ~,
\nonumber \\
v_{R} &\equiv& \overrightarrow{v} \cdot \overrightarrow{e}_{R} = \sqrt{\frac{G~M_{\odot} \left ( 1 - \beta \right )}{p_{\beta}}} ~
e_{\beta} ~ \sin{f_{\beta}} ~,
\nonumber \\
v_{T} &\equiv& \overrightarrow{v} \cdot \overrightarrow{e}_{T} =
\sqrt{\frac{G~M_{\odot} \left ( 1 - \beta \right )}{p_{\beta}}} ~ \left ( 1 + e_{\beta} ~\cos{f_{\beta}} \right ) ~,
\nonumber \\
r &=& p_{\beta} / \left ( 1 + e_{\beta} ~\cos{f_{\beta}} \right ) ~,
\end{eqnarray}
where $p_{\beta}$ $=$ $a_{\beta} (1 - e_{\beta}^{2})$ is the semi-latus rectum, $a_{\beta}$ the semi-major axis, $e_{\beta}$ the osculating
eccentricity of the orbit and $f_{\beta}$ is the true anomaly. Inserting Eqs. (\ref{celmech1}) into perturbation equations
of celestial mechanics for $a_{\beta}$, $e_{\beta}$ and longitude of perihelion $\omega_{\beta}$,
\begin{eqnarray}\label{celmech2}
\frac{d a_{\beta}}{dt} &=& \frac{2 a_{\beta}}{1 - e_{\beta}^{2}} ~
\sqrt{\frac{p_{\beta}}{G~M_{\odot} \left ( 1 - \beta \right )}} \times
\nonumber \\
& & \left [ F_{\beta R} ~e_{\beta} ~ \sin{f_{\beta}} +
F_{\beta T} \left ( 1 + e_{\beta} ~\cos{f_{\beta}} \right ) \right ] ~,
\nonumber \\
\frac{d e_{\beta}}{dt} &=& \sqrt{\frac{p_{\beta}}{G~M_{\odot} \left ( 1 - \beta \right )}} \times
\nonumber \\
& & \left [ F_{\beta R} \sin{f_{\beta}} +
F_{\beta T} \left ( \cos{f_{\beta}} + \frac{e_{\beta} + \cos{f_{\beta}}}{1 + e_{\beta} \cos{f_{\beta}}} \right ) \right ] ~,
\nonumber \\
\frac{d \omega_{\beta}}{dt} &=& -~\frac{1}{e_{\beta}} ~
\sqrt{\frac{p_{\beta}}{G~M_{\odot} \left ( 1 - \beta \right )}} \times
\nonumber \\
& & \left [ F_{\beta R} ~ \cos{f_{\beta}} -
F_{\beta T} \frac{2 + e_{\beta} ~\cos{f_{\beta}}}{1 + e_{\beta} ~\cos{f_{\beta}}}  ~\sin{f_{\beta}} \right ] ~,
\nonumber \\
p_{\beta} &=& a_{\beta} \left ( 1 - e_{\beta}^{2} \right ) ~,
\end{eqnarray}
neglecting change of the mass of the IDP and its optical properties, and, using time averaging (approximation of the
exact averaging given by Eq. 103 in Kla\v{c}ka 1992a)
\begin{eqnarray}\label{celmech3}
\langle g \rangle = \frac{1}{a_{\beta}^{2} \sqrt{1 - e_{\beta}^{2}}} ~\frac{1}{2 \pi} ~\int_{0}^{2 \pi} g \left ( f_{\beta} \right )
r^{2} d f_{\beta} ~,
\end{eqnarray}
one can easily obtain the secular evolution of $a_{\beta}$ and $e_{\beta}$ ( $g$ $\rightarrow$ $d a_{\beta} / dt$, $d e_{\beta} / dt$
and omitting the brackets $\langle \rangle$):
\begin{eqnarray}\label{celmech4}
\frac{d a_{\beta}}{dt} &=& - ~ \beta ~\frac{G~M_{\odot}}{c} \times
\nonumber \\
& & \frac{2 \left ( 1 + \eta_{2} / \overline{Q}'_{pr} \right )
+ \left ( 3 + \eta_{1} / \overline{Q}'_{pr} + 2 \eta_{2} / \overline{Q}'_{pr} \right ) e_{\beta}^{2}}{a_{\beta} \left ( 1 - e_{\beta}^{2} \right )^{3/2}}
\nonumber \\
& & +~ 2 ~\gamma_{T} ~\frac{\eta_{2}}{\overline{Q} ~'_{pr}} ~\beta ~\frac{G~M_{\odot}}{c}
\frac{u / \sqrt{G~M_{\odot} \left ( 1 - \beta \right ) / p_{\beta}}}{
a_{\beta} \left ( 1 - e_{\beta}^{2} \right )^{3/2}}
\nonumber \\
& & -~ 3 ~\gamma_{T} ~\beta ~\frac{G~M_{\odot}}{c}
\frac{\sqrt{G~M_{\odot} \left ( 1 - \beta \right ) / p_{\beta}} ~/~ u}{
a_{\beta} \left ( 1 - e_{\beta}^{2} \right )^{3/2}}
\nonumber \\
& & \times \left [ \frac{1}{2} ~\frac{\eta_{3}}{\overline{Q} ~'_{pr}} ~e_{\beta}^{2}
- \frac{\eta_{1}}{\overline{Q} ~'_{pr}} \left ( 1 + 2 e_{\beta}^{2} \right ) \right ] ~,
\nonumber \\
\frac{d e_{\beta}}{dt} &=&  - ~\frac{5 + \eta_{1} / \overline{Q}'_{pr} + 4~\eta_{2} / \overline{Q}'_{pr}}{2}
~\beta ~\frac{G~M_{\odot}}{c}
\nonumber \\
& & \times \frac{e_{\beta}}{a_{\beta}^{2} ~\sqrt{1 - e_{\beta}^{2}}}
\nonumber \\
& & +~ \gamma_{T} ~\frac{\eta_{2}}{\overline{Q} ~'_{pr}} ~ \beta ~\frac{G~M_{\odot}}{c}
\frac{u}{\sqrt{G~M_{\odot} \left ( 1 - \beta \right ) / p_{\beta}}}
\nonumber \\
& & \times \frac{1 - \sqrt{1 - e_{\beta}^{2}}}{a_{\beta}^{2} ~e_{\beta} ~\sqrt{1 - e_{\beta}^{2}}}
\nonumber \\
& & +~ \gamma_{T}  ~\beta ~\frac{G M_{\odot}}{c}
\frac{\sqrt{G M_{\odot} \left ( 1 - \beta \right ) / p_{\beta}} /u}{a_{\beta}^{2} \sqrt{1 - e_{\beta}^{2}}} \times X_{e \beta} ~,
\nonumber \\
X_{e \beta} &=& \left ( \frac{15}{4} \frac{\eta_{1}}{\overline{Q}'_{pr}} - \frac{\eta_{3}}{\overline{Q}'_{pr}} \right ) e_{\beta}
\nonumber \\
& & +~ \frac{\eta_{1} - \eta_{3}}{4 ~\overline{Q}'_{pr}} \left [
1 - 2 \frac{\left ( 1 - \sqrt{1 - e_{\beta}^{2}} \right ) \left ( 1 - e_{\beta}^{2} \right )}{e_{\beta}^{2}} \right ] \frac{1}{e_{\beta}} ~,
\nonumber \\
\frac{d \omega_{\beta}}{dt} &=&  -~ \frac{\eta_{1}}{\overline{Q}'_{pr}}  ~\beta \frac{G M_{\odot}}{c}
\frac{1}{a_{\beta}^{2} ~\sqrt{1 - e_{\beta}^{2}}}
\nonumber \\
& & \times \left [ \gamma_{T} \frac{1 - \sqrt{1 - e_{\beta}^{2}}}{e_{\beta}^{2}} - \frac{1}{2}
\frac{\sqrt{G M_{\odot} \left ( 1 - \beta \right ) / p_{\beta}}}{u} \right ] ~,
\nonumber \\
p_{\beta} &=& a_{\beta} \left ( 1 - e_{\beta}^{2} \right ) ~.
\end{eqnarray}
Initial conditions can be found in Sec. 6.1 in Kla\v{c}ka (2004) and will not be repeated here.

Eqs. (\ref{celmech4}) show that a systematic secular decrease of semi-major axis and eccentricity exists, if the particle is not ejected from the
Solar System due to the radiation pressure, for the case $\gamma_{T}$ $\equiv$ 0. However, the case $\gamma_{T}$ $\ne$ 0 for prograde
orbits is more interesting. The first of Eqs. (\ref{celmech4}) enables also $d a_{\beta} / dt$ $>$ 0:
\begin{eqnarray}\label{celmech4a1}
a_{\beta} [AU] &>& \left ( 1 - \beta \right )
\left ( \frac{v_{E} / u}{2 ~\gamma_{T} ~\eta_{2} / \overline{Q} ~'_{pr}} \right )^{2} \times \frac{X_{a \beta 0}^{2}}{1 - e_{\beta}^{2}} ~,
\nonumber \\
X_{a \beta 0} &\equiv& 2 \left ( 1 + \eta_{2} / \overline{Q}'_{pr} \right )
\nonumber \\
& & + \left ( 3 + \eta_{1} / \overline{Q}'_{pr} + 2 \eta_{2} / \overline{Q}'_{pr} \right ) e_{\beta}^{2}  ~,
\nonumber \\
\frac{d a_{\beta}}{dt} &>& 0 ~,
\end{eqnarray}
or, numerically,
\begin{eqnarray}\label{celmech4a2}
a_{\beta} [AU] &>& \frac{1 - \beta}{4.834} ~ \frac{\left ( 4.8 + 6.9~ e_{\beta}^{2} \right )^{2}}{1 - e_{\beta}^{2}} ~,
\nonumber \\
& & \frac{d a_{\beta}}{dt} > 0 ~,
\end{eqnarray}
if Eqs. (\ref{eta-1-2}) are used for $\eta_{1}$, $\eta_{2}$, $u$ $=$ 450 $km~s^{-1}$,
$v_{E}$ $=$ 29.8 $km~s^{-1}$, $\gamma_{T}$ $=$ 0.052, and, $\overline{Q} ~'_{pr}$ $=$ 1.
As an example we can mention that $d a_{\beta} / dt$ $>$ 0 can be fulfilled already for
$a_{\beta}$ $>$ 4.8 $AU$ for $e_{\beta}$ $\rightarrow$ 0.
Similarly, the second of Eqs. (\ref{celmech4}) enables also $d e_{\beta} / dt$ $>$ 0:
\begin{eqnarray}\label{celmech4e1}
a_{\beta} [AU] &>& \left ( 1 - \beta \right )
\left ( \frac{v_{E} / u}{2 ~\gamma_{T} ~\eta_{2} / \overline{Q} ~'_{pr}} \right )^{2}
\nonumber \\
& & \times  \left [ \frac{\left ( 5 + \eta_{1} / \overline{Q} ~'_{pr} + 4~\eta_{2} / \overline{Q} ~'_{pr} \right ) e_{\beta}^{2}}{
\sqrt{1 - e_{\beta}^{2}} - 1 + e_{\beta}^{2}} \right ]^{2} ~,
\nonumber \\
\frac{d e_{\beta}}{dt} &>& 0 ~,
\end{eqnarray}
or, numerically,
\begin{eqnarray}\label{celmech4e2}
a_{\beta} [AU] &>& \frac{1 - \beta}{4.834} ~ \left ( \frac{11.7~ e_{\beta}^{2}}{\sqrt{1 - e_{\beta}^{2}} - 1 + e_{\beta}^{2}}
\right )^{2} ~,
\nonumber \\
& & \frac{d e_{\beta}}{dt} > 0 ~,
\end{eqnarray}
if Eqs. (\ref{eta-1-2}) are used for $\eta_{1}$, $\eta_{2}$, $u$ $=$ 450 $km~s^{-1}$,
$v_{E}$ $=$ 29.8 $km~s^{-1}$, $\gamma_{T}$ $=$ 0.052, and, $\overline{Q} ~'_{pr}$ $=$ 1.
The case  $d e_{\beta} / dt$ $>$ 0 does not realize since, e.g., $a_{\beta}$ $>$ 113.2 $AU$ for $e_{\beta}$ $\rightarrow$ 0
and $a_{\beta}$ $>$ 237.9 $AU$ for  $e_{\beta}$ $=$ 0.85.
The solar wind does not act at distances greater than about (100-150) $AU$ and, moreover,
the action of fast interstellar gas flow plays an important role at distances greater than about 20 $AU$
(P\'{a}stor et al. 2011).

Eqs. (\ref{celmech4}) can be partially analytically integrated for the case $\gamma_{T}$ $\equiv$ 0.
The semi-latus rectum $p_{\beta}$ $=$ $a_{\beta} (1 - e_{\beta}^{2})$
and Eqs. (\ref{celmech4}) yield
\begin{eqnarray}\label{celmech5}
\frac{d p_{\beta}}{dt} &=& - ~2 ~ \beta ~\frac{G~M_{\odot}}{c} ~ \left ( 1 + \frac{\eta_{2}}{\overline{Q}'_{pr}} \right ) ~
\frac{\left ( 1 - e_{\beta}^{2} \right )^{3/2}}{p_{\beta}} ~,
\nonumber \\
\frac{d e_{\beta}}{dt} &=&  - ~\frac{5 + \eta_{1} / \overline{Q}'_{pr} + 4 \eta_{2} / \overline{Q}'_{pr}}{2}
\nonumber \\
& & \times \beta ~\frac{G M_{\odot}}{c} ~\frac{e_{\beta} \left ( 1 - e_{\beta}^{2} \right )^{3/2}}{p_{\beta}^{2}} ~.
\end{eqnarray}
It can be easily verified that Eqs. (\ref{celmech5}) give
\begin{eqnarray}\label{celmech6}
p_{\beta} &=& p_{\beta in} ~ \left ( \frac{e_{\beta}}{e_{\beta in}} \right )^{\alpha_{w}} ~,
\nonumber \\
\alpha_{w} &=& \frac{4 \left ( 1 + \eta_{2} / \overline{Q} ~'_{pr} \right )}{5 + \eta_{1} / \overline{Q} ~'_{pr} + 4 ~\eta_{2} / \overline{Q} ~'_{pr}} ~,
\end{eqnarray}
where $p_{\beta in}$ and $e_{\beta in}$ are initial values of the semi-latus rectum and the eccentricity.

Eqs. (\ref{celmech6}) and the second of Eqs. (\ref{celmech5}) yield
\begin{eqnarray}\label{celmech7}
\frac{d e_{\beta}}{dt} &=&  - ~\frac{5 + \eta_{1} / \overline{Q} ~'_{pr} + 4~\eta_{2} / \overline{Q} ~'_{pr}}{2}
~\beta ~\frac{G~M_{\odot}}{c}
\nonumber \\
& & \times \frac{\left ( e_{\beta in} \right )^{2 \alpha_{w}}}{\left ( p_{\beta in} \right )^{2}}
\frac{\left ( 1 - e_{\beta}^{2} \right )^{3/2}}{e_{\beta}^{2 \alpha_{w} -~ 1}} ~.
\end{eqnarray}
Eq. (\ref{celmech7}) shows that the eccentricity is a decreasing function of time and the value $e_{\beta}$
$=$ 0 occurs at a finite time. Eqs. (\ref{celmech6}) show that $p_{\beta}$ $=$ $a_{\beta}$ $=$ 0 at the same finite time.
The time of spiralling toward the Sun from some initial values
$e_{\beta in}$ and $p_{\beta in}$ to values $e_{\beta}$ and $p_{\beta}$ is
\begin{eqnarray}\label{celmech8}
\tau ( e_{\beta in} ; e_{\beta} ; \eta_{1} ; \eta_{2} ) &=& -~ \frac{2}{5 + \eta_{1} / \overline{Q} ~'_{pr} + 4~\eta_{2} / \overline{Q} ~'_{pr}}
\nonumber \\
& & \times \left ( \beta ~\frac{G~M_{\odot}}{c} \right )^{-1}
\frac{\left ( p_{\beta in} \right )^{2}}{\left ( e_{\beta in} \right )^{2 \alpha_{w}}}
\nonumber \\
& & \times \int_{e_{\beta in}}^{e_{\beta}}
\frac{x^{2 \alpha_{w} -~ 1}}{\left ( 1 - x^{2} \right )^{3/2}} ~\mbox{d}x ~,
\nonumber \\
p_{\beta in} &=& a_{\beta in} \left ( 1 - e_{\beta in}^{2} \right ) ~,
\nonumber \\
\alpha_{w} &=& \frac{4 \left ( 1 + \eta_{2} / \overline{Q} ~'_{pr} \right )}{5 + \eta_{1} / \overline{Q} ~'_{pr} + 4 ~\eta_{2} / \overline{Q} ~'_{pr}} ~.
\end{eqnarray}

The conventional case can be obtained from Eqs. (\ref{celmech6}) and (\ref{celmech8}):
\begin{eqnarray}\label{celmech8-convention}
\tau ( e_{\beta in} ; e_{\beta} ; \eta_{0} ; \eta_{0} ) &=& -~ \frac{2}{5}
\left [ \beta  \left ( 1 + \frac{\eta_{0}}{\overline{Q} ~'_{pr}} \right ) ~\frac{G~M_{\odot}}{c} \right ]^{-1}
\nonumber \\
& & \times ~\frac{\left ( p_{\beta in} \right )^{2}}{\left ( e_{\beta in} \right )^{8/5}} ~
\int_{e_{\beta in}}^{e_{\beta}} \frac{x^{3/5}}{(1 - x^{2} )^{3/2}} ~\mbox{d}x ~,
\nonumber \\
p_{\beta} &=& p_{\beta in} \left ( \frac{e_{\beta}}{e_{\beta in}} \right )^{4/5}  ~,
\nonumber \\
\eta_{0} &=& 0.3 ~,
\end{eqnarray}
where the conventional value of $\eta_{0}$ is added. Eqs. (\ref{celmech8})-(\ref{celmech8-convention}) yield
\begin{eqnarray}\label{celmech8-comparison}
\frac{\tau ( e_{\beta in} ; e_{\beta} ; \eta_{1} ; \eta_{2} )}{\tau ( e_{\beta in} ; e_{\beta} ; \eta_{0} ; \eta_{0} )} &=&
\frac{5 \left ( 1 + \eta_{0} / \overline{Q} ~'_{pr} \right )}{5 + \left ( \eta_{1} + 4~\eta_{2} \right ) / \overline{Q} ~'_{pr}}
\nonumber \\
& & \times \left ( e_{\beta in} \right )^{8/5 - ~2 \alpha_{w}} ~
\frac{X_{\tau 12}}{X_{\tau 00}} ~,
\nonumber \\
X_{\tau 12} &\equiv& \int_{e_{\beta in}}^{e_{\beta}}
\frac{x^{2 \alpha_{w} -~ 1}}{\left ( 1 - x^{2} \right )^{3/2}} ~\mbox{d}x ~,
\nonumber \\
X_{\tau 00} &\equiv& \int_{e_{\beta in}}^{e_{\beta}}
\frac{x^{3/5}}{\left ( 1 - x^{2} \right )^{3/2}} ~\mbox{d}x ~.
\end{eqnarray}
Eqs. (\ref{eta-1-2}), (\ref{celmech6}), (\ref{celmech8-convention}) and (\ref{celmech8-comparison}) give the following result:
the real time of inspiralling toward the Sun is (0.54-0.56)-multiple of the conventionally considered time
for the relative time of inspiralling toward the Sun, from some $e_{\beta in}$ to $e_{\beta}$ $=$ 0, see Table 6.

\begin{table}
\begin{center}
\begin{tabular}{|c|c|}
\hline
$e_{\beta in}$ & $\tau_{real} / \tau_{conv}$  \\
\hline
0.001  & 0.5417  \\
0.010  & 0.5417  \\
0.020  & 0.5417  \\
0.050  & 0.5417  \\
0.100  & 0.5417  \\
0.200  & 0.5419  \\
0.250  & 0.5420  \\
0.500  & 0.5431  \\
0.750  & 0.5454  \\
0.800  & 0.5462  \\
0.850  & 0.5471  \\
0.900  & 0.5484  \\
0.950  & 0.5502  \\
0.990  & 0.5529  \\
0.999  & 0.5547  \\
\hline
\end{tabular}
\caption{The ratio $\tau_{real} / \tau_{conv}$ $\equiv$
$\tau ( e_{\beta in} ; 0 ; \eta_{1} ; \eta_{2} ) / \tau ( e_{\beta in} ; 0 ; \eta_{0} ; \eta_{0} )$
of the times of inspiralling of a spherical IDP toward the Sun. Real time is compared with
the conventional approach to the time of inspiralling. Various initial eccentricities $e_{\beta in}$ are considered,
$\overline{Q} ~'_{pr} = 1$, $\eta_{1}$ $=$ 1.1, $\eta_{2}$ $=$ 1.4, $\eta_{0}$ $=$ 0.3.}
\end{center}
\label{tab:6}
\end{table}

Eqs. (\ref{celmech8}) can be rewritten to the form
\begin{eqnarray}\label{celmech8num}
\tau ( e_{\beta in} ; e_{\beta} = 0 ; \eta_{1} ; \eta_{2} ) &=&
\frac{2 \left ( \beta ~G~M_{\odot} / c \right )^{-1}}{5 + \eta_{1} / \overline{Q} ~'_{pr} + 4~\eta_{2} / \overline{Q} ~'_{pr}}
\nonumber \\
& & \times \left ( p_{\beta in} \right )^{2} ~ \tau_{TAB} ( e_{\beta in} ) ~,
\nonumber \\
\eta_{1} &=& 1.1 ~, ~~ \eta_{2} = 1.4 ~,
\nonumber \\
p_{\beta in} &=& a_{\beta in} \left ( 1 - e_{\beta in}^{2} \right ) ~,
\end{eqnarray}
where the function $\tau_{TAB} ( e_{\beta in} )$ is represented by the data collected in Table 7.

\begin{table}
\begin{center}
\begin{tabular}{|r|r|}
\hline
$e_{\beta in}$ & $\tau_{TAB} ( e_{\beta in} )$    \\
\hline
0.001  &  0.6094  \\
0.250  &  0.6365  \\
0.500  &  0.7389  \\
0.750  &  1.0504  \\
0.800  &  1.1893  \\
0.850  &  1.4005  \\
0.900  &  1.7683  \\
0.950  &  2.6327  \\
0.990  &  6.4436  \\
0.999  & 21.6686  \\
\hline
\end{tabular}
\caption{Values of the function $\tau_{TAB} ( e_{\beta in} )$ for various initial eccentricities $e_{\beta in}$, see
Eqs. (\ref{celmech8num}) describing time of inspiralling toward the Sun due to the action of the solar radiation,
both electromagnetic and corpuscular.}
\end{center}
\label{tab:7}
\end{table}

\subsection{Stars}
Our results may be applied to dust dynamics in disks around stars with stellar winds
(see, e.g., Strubbe and Chiang 2006, Plavchan et al. 2009).

\subsubsection{The time of spiralling toward the central star from an initial orbit}
On the basis of Eqs. (\ref{beta-eta-kappa}) and (\ref{celmech8}) we can write for
the time of spiralling toward a central star from some initial values
$e_{\beta in}$ and $p_{\beta in}$ to values $e_{\beta}$ and $p_{\beta}$:
\begin{eqnarray}\label{celmech8-star}
\tau ( e_{\beta in} ; e_{\beta} ; \eta_{1} ; \eta_{2} ) &=& -~ \frac{2}{5 ~+~ \eta_{1} / \overline{Q} ~'_{pr} ~+~ 4~\eta_{2} / \overline{Q} ~'_{pr}}
\nonumber \\
& & \times \left ( \beta ~\frac{G~M_{\star}}{c} \right )^{-1}
\frac{\left ( p_{\beta in} \right )^{2}}{\left ( e_{\beta in} \right )^{2 \alpha_{w}}}
\nonumber \\
& & \times \int_{e_{\beta in}}^{e_{\beta}}
\frac{x^{2 \alpha_{w} -~ 1}}{\left ( 1 - x^{2} \right )^{3/2}} ~\mbox{d}x ~,
\nonumber \\
p_{\beta in} &=& a_{\beta in} \left ( 1 - e_{\beta in}^{2} \right ) ~,
\nonumber \\
\alpha_{w} &=& \frac{4 \left ( 1 + \eta_{2} / \overline{Q} ~'_{pr} \right )}{5 + \eta_{1} / \overline{Q} ~'_{pr} + 4 ~\eta_{2} / \overline{Q} ~'_{pr}} ~,
\end{eqnarray}
where
\begin{eqnarray}\label{beta-eta-kappa-star}
\beta &=& \frac{L_{\star} ~R^{2} ~\overline{Q} ~'_{pr}}{4 G M_{\star} m c} ~,
\nonumber \\
\eta_{j} &=& \frac{c^{2} ~m_{p} ~u~ n_{p} (r_{0})}{2 ~S_{\star 0}} ~\tilde{\eta}_{j} \equiv
\frac{c^{2} ~m_{p} ~u~ n_{p}(r_{0})}{2 \left [ L_{\star} / \left ( 4 \pi ~r_{0}^{2} \right ) \right ]}
~\tilde{\eta}_{j}  ~,
\nonumber \\
& & ~~ j = 1, 2, 3 ~.
\end{eqnarray}
Here $L_{\star}$ is the rate of energy outflow from the star, the stellar luminosity, $R$ is the radius of the IDP, $m$ its mass,
$\overline{Q} ~'_{pr}$ is the dimensionless efficiency factor of the radiation pressure averaged over the solar spectrum,
$G$ is the gravitational constant, $M_{\star}$ is the mass of the star, $c$ is the speed of light in vacuum
$m_{p}$ is the proton mass, $n_{p}$ is the proton concentration in the stellar wind and
$S_{\star 0}$ denotes the stellar electromagnetic flux at a distance $r_{0}$ from the star
($S_{\odot 0}$ $\equiv$ $S_{0}$ $=$ 1.366 $\times$ $10^{3}$ $W~m^{-2}$ is the solar electromagnetic flux).
As for the homogeneous spherical IDP we have $m$ $=$ ($4 \pi / 3$) $\rho$ $R^{3}$, where $\rho$ is the mass density of the IDP.

Eq. (\ref{celmech8-star}) yields
\begin{eqnarray}\label{celmech8-star-sw}
\tau_{SW} ( e_{\beta in} ; e_{\beta} ; \eta_{1} ; \eta_{2} ) &=& -~ \frac{2}{\eta_{1} / \overline{Q} ~'_{pr} + 4~\eta_{2} / \overline{Q} ~'_{pr}}
\nonumber \\
& & \times \left ( \beta ~\frac{G~M_{\star}}{c} \right )^{-1} ~\frac{\left ( p_{\beta in} \right )^{2}}{\left ( e_{\beta in} \right )^{2 \alpha_{w0}}}
\nonumber \\
& & \times ~\int_{e_{\beta in}}^{e_{\beta}}
\frac{x^{2 \alpha_{w0} -~ 1}}{\left ( 1 - x^{2} \right )^{3/2}} ~\mbox{d}x ~,
\nonumber \\
p_{\beta in} &=& a_{\beta in} \left ( 1 - e_{\beta in}^{2} \right ) ~,
\nonumber \\
\alpha_{w0} &=& \frac{4 ~\eta_{2}}{\eta_{1} + 4 ~\eta_{2}} ~,
\end{eqnarray}
for the stellar wind, i.e., corpuscular radiation, and,
\begin{eqnarray}\label{celmech8-star-elmg}
\tau_{E} ( e_{\beta in} ; e_{\beta} ) &=& -~ \frac{2}{5} ~\left ( \beta ~\frac{G~M_{\star}}{c} \right )^{-1}
\nonumber \\
& & \times ~\frac{\left ( p_{\beta in} \right )^{2}}{\left ( e_{\beta in} \right )^{8 / 5}} ~
\int_{e_{\beta in}}^{e_{\beta}}
\frac{x^{3 / 5}}{\left ( 1 - x^{2} \right )^{3/2}} ~\mbox{d}x ~,
\nonumber \\
p_{\beta in} &=& a_{\beta in} \left ( 1 - e_{\beta in}^{2} \right ) ~,
\end{eqnarray}
for the electromagnetic radiation of the star.

Eqs. (\ref{celmech8-star-sw}) and (\ref{celmech8-star-elmg}) lead to
\begin{eqnarray}\label{celmech8-star-ratio}
\frac{\tau_{SW} ( e_{\beta in} ; e_{\beta} ; \eta_{1} ; \eta_{2} )}{\tau_{E} ( e_{\beta in} ; e_{\beta} )}
&=& \frac{5}{\eta_{1} / \overline{Q} '_{pr} + 4 \eta_{2} / \overline{Q} '_{pr}}
\frac{\left ( e_{\beta in} \right )^{8 / 5}}{\left ( e_{\beta in} \right )^{2 \alpha_{w0}}}
\nonumber \\
& & \times ~\int_{e_{\beta in}}^{e_{\beta}}
\frac{x^{2 \alpha_{w0} -~ 1}}{\left ( 1 - x^{2} \right )^{3/2}} ~\mbox{d}x
\nonumber \\
& & \times \left \{ \int_{e_{\beta in}}^{e_{\beta}}
\frac{x^{3 / 5}}{\left ( 1 - x^{2} \right )^{3/2}} ~\mbox{d}x \right \} ^{-1} ~,
\nonumber \\
\alpha_{w0} &=& \frac{4 ~\eta_{2}}{\eta_{1} + 4 ~\eta_{2}}  ~.
\end{eqnarray}

Eq. (\ref{celmech8-star-ratio}) yields for the near circular orbits
\begin{eqnarray}\label{celmech8-star-ratio-e0-1}
\frac{\tau_{SW} ( e_{\beta in} ; e_{\beta} ; \eta_{1} ; \eta_{2} )}{\tau_{E} ( e_{\beta in} ; e_{\beta} )}
&=& \frac{5}{\eta_{1} / \overline{Q} ~'_{pr} + 4~\eta_{2} / \overline{Q} ~'_{pr}} ~ X_{SWE} ~,
\nonumber \\
X_{SWE} &=& \frac{1}{\left ( e_{\beta in} \right )^{2 \alpha_{w0}}} \int_{e_{\beta in}}^{e_{\beta}} x^{2 \alpha_{w0} - 1} \mbox{d}x
\nonumber \\
& & \times \left \{ \frac{1}{\left ( e_{\beta in} \right )^{8 / 5}} \int_{e_{\beta in}}^{e_{\beta}} x^{3 / 5} \mbox{d}x \right \} ^{-1}
\nonumber \\
&=& \frac{4}{5~\alpha_{w0}} ~
\frac{\left ( e_{\beta} / e_{\beta in} \right )^{2 \alpha_{w0}} - 1}{\left ( e_{\beta} / e_{\beta in} \right )^{8 / 5} - 1} ~,
\nonumber \\
\alpha_{w0} &=& \frac{4 ~\eta_{2}}{\eta_{1} + 4 ~\eta_{2}} ~,
\nonumber \\
e_{\beta in} &\ll& 1 ~,
\end{eqnarray}
or, shortly,
\begin{eqnarray}\label{celmech8-star-ratio-e0-2}
\frac{\tau_{SW}}{\tau_{E}} &=& \frac{\overline{Q} ~'_{pr}}{\eta_{2}} ~,
\nonumber \\
\frac{\tau_{SW}}{\tau_{E}} &\equiv& \frac{\tau_{SW} ( e_{\beta in} \ll 1; e_{\beta} ; \eta_{1} ; \eta_{2} )}{\tau_{E} ( e_{\beta in} \ll 1; e_{\beta} )} ~.
\end{eqnarray}
More straightforward approach uses the first of Eqs. (\ref{celmech4}), when $\gamma_{T}$ $=$ 0 and, formally, $e_{\beta}$ $=$ 0.
Similarly, Eqs. (\ref{celmech8-star}),
(\ref{celmech8-star-sw}) and (\ref{celmech8-star-elmg}), give
\begin{eqnarray}\label{celmech8-star-ratio-e0-2-rest}
\frac{\tau_{SW}}{\tau} &=& 1 ~+~ \left ( \frac{\eta_{2}}{\overline{Q} ~'_{pr}} \right ) ^{-1} ~,
\nonumber \\
\frac{\tau_{E}}{\tau} &=& 1 ~+~ \frac{\eta_{2}}{\overline{Q} ~'_{pr}}  ~,
\nonumber \\
\frac{1}{\tau} &=& \frac{1}{\tau_{E}} ~+~ \frac{1}{\tau_{SW}} ~,
\nonumber \\
e_{\beta in} &\ll& 1 ~.
\end{eqnarray}
Explicitly,
\begin{eqnarray}\label{celmech8-star-complete}
\tau ( e_{\beta in} \ll 1 ; e_{\beta} ; \eta_{1} ; \eta_{2} ) &=&  \frac{1}{4}  ~ \frac{1}{1 ~+~ \eta_{2} / \overline{Q} ~'_{pr}}
\nonumber \\
& & \times \left ( \beta \frac{G M_{\star}}{c} \right )^{-1} \left ( a_{\beta in} \right )^{2} ~,
\nonumber \\
\tau_{SW} ( e_{\beta in} \ll 1 ; e_{\beta} ; \eta_{1} ; \eta_{2} ) &=&  \frac{1}{4} \frac{\overline{Q}~'_{pr}}{\eta_{2}}
\nonumber \\
& & \times \left ( \beta \frac{G M_{\star}}{c} \right )^{-1} \left ( a_{\beta in} \right )^{2} ~,
\nonumber \\
\tau_{E} ( e_{\beta in} \ll 1 ; e_{\beta} ) &=&  \frac{1}{4}  ~
\left ( \beta ~\frac{G~M_{\star}}{c} \right )^{-1}
\nonumber \\
& & \times \left ( a_{\beta in} \right )^{2} ~.
\end{eqnarray}

We can write, as a good approximation,
\begin{eqnarray}\label{celmech8-star-ratio-e0-3}
\frac{\tau_{SW}}{\tau_{E}} &\doteq&
\frac{\overrightarrow{F}_{P-R} \cdot \overrightarrow{e}_{T}}{\overrightarrow{F}_{SW} \cdot \overrightarrow{e}_{T}} ~,
\nonumber \\
\frac{\tau_{SW}}{\tau_{E}} &\doteq&
\frac{u}{c} ~\frac{\overrightarrow{F}_{P-R} \cdot \overrightarrow{e}_{R}}{\overrightarrow{F}_{SW} \cdot \overrightarrow{e}_{R}}
~\left \{  1 - \left ( 1 + \frac{\eta_{1}}{\eta_{2}} \right ) ~\frac{\overrightarrow{v} \cdot \overrightarrow{e}_{R}}{u} \right \}
\nonumber \\
&\doteq& \frac{u}{c} ~\frac{\overrightarrow{F}_{P-R} \cdot \overrightarrow{e}_{R}}{\overrightarrow{F}_{SW} \cdot \overrightarrow{e}_{R}} ~,
\end{eqnarray}
where $\overrightarrow{F}_{P-R}$ is the Poynting-Robertson force (electromagnetic radiation pressure force)
and $\overrightarrow{F}_{SW}$ is the stellar wind force (corpuscular radiation pressure force):
\begin{eqnarray}\label{accel-P-R}
\overrightarrow{F}_{P-R} &\doteq& \beta ~\frac{G~M_{\star} ~m}{r^{2}}
\nonumber \\
& & \times \left [ \left ( 1 ~-~ \frac{\overrightarrow{v} \cdot \overrightarrow{e}_{R}}{c}  \right ) \overrightarrow{e}_{R}
~-~ \frac{\overrightarrow{v}}{c} \right ] ~,
\end{eqnarray}
\begin{eqnarray}\label{accel-SW}
\overrightarrow{F}_{SW} &\doteq& \beta ~\frac{G~M_{\star} ~m}{r^{2}}
\nonumber \\
& & \times \left ( \frac{\eta_{2}}{\overline{Q} ~'_{pr}} ~ \frac{u}{c}
~-~ \frac{\eta_{1}}{\overline{Q} ~'_{pr}} ~
\frac{\overrightarrow{v} \cdot \overrightarrow{e}_{R}}{c}  \right ) \overrightarrow{e}_{R}
\nonumber \\
& & -~ \beta ~\frac{G~M_{\star} ~m}{r^{2}} \frac{\eta_{2}}{\overline{Q} ~'_{pr}} ~ \frac{\overrightarrow{v}}{c}~,
\end{eqnarray}
where the change of test particle mass $m$, due to the action of the stellar wind, is neglected.
We remind that the electromagnetic radiation pressure force contains both the non-velocity and velocity terms of the body moving with respect
to the central star, the source of the electromagnetic radiation; the same holds for the corpuscular radiation pressure force.
As for the difference between the results holding for the $\kappa$ and the Maxwell-Boltzmann distributions, one can compare
Eq. (\ref{accel-SW}) with Eq. (30) in Kla\v{c}ka (2013). The forces presented by
Eqs. (\ref{accel-P-R})-(\ref{accel-SW}) are drag forces.

Eq. (\ref{celmech8-star-ratio-e0-2}) can be rewritten to the form
\begin{eqnarray}\label{celmech8-star-ratio-e0-4}
\frac{\tau_{SW}}{\tau_{E}} &=& \frac{\overline{Q} ~'_{pr}}{\eta_{\odot 2}}
~\frac{\tilde{\eta}_{\odot 2}}{\tilde{\eta}_{\star 2}}
~\frac{L_{\star}}{L_{\odot}}
~\frac{\dot{M}_{\odot}}{\dot{M}_{\star}} ~,
\end{eqnarray}
if also the second of Eqs. (\ref{beta-eta-kappa-star}) is used.
The result represented by Eq. (\ref{celmech8-star-ratio-e0-4}) differs from that presented by Plavchan et. al (2009 - Eq. A10), and, also
from Eqs. (31)-(32) by Kla\v{c}ka (2013).

\subsubsection{The ratio of repulsive to gravitational forces}
As an illustration of the importance of our new results,
we will shortly discuss the ratio of repulsive to gravitational forces. Defining the parameter
\begin{eqnarray}\label{stars-1}
\beta_{ESWR} &\equiv& \frac{F_{elmg} + F_{wind}}{F_{grav}}
\nonumber \\
&=& \frac{3}{16~ \pi} ~\frac{L_{\star} ~P_{ESWR}}{G ~M_{\star} ~c ~\rho ~R} ~,
\end{eqnarray}
where the dimensionless factor
\begin{eqnarray}\label{stars-2}
P_{ESWR} &=& \overline{Q} ~'_{pr} + Q_{wind}~ \frac{\dot{M}_{\star} ~u ~c}{L_{\star}}
\end{eqnarray}
measures the extent to which the pressure exerted by the radial wind dominates electromagnetic radiation pressure,
see, e.g., Eqs. (5)-(7) in Strubbe and Chiang (2006). We remind that $L_{\star}$ is the luminosity of the star, $M_{\star}$ its mass
and $\dot{M}_{\star}$ is the amount of stellar wind mass ejected by the star per unit time. The quantity $Q_{wind}$
is the dimensionless cross section the dust grain presents to wind pressure. Comparing with our physical
approach, we can write
\begin{eqnarray}\label{stars-3}
Q_{wind} &=& \eta_{2} ~\frac{L_{\star}}{\dot{M}_{\star} ~c^{2}} ~.
\end{eqnarray}

Eqs. (\ref{stars-1})-(\ref{stars-2}) can be rewritten to the form
\begin{eqnarray}\label{stars-1-new}
\beta_{ESWR} &=& \beta_{E} + \frac{3}{16~ \pi} ~Q_{wind} ~\frac{\dot{M}_{\star}}{M_{\star}} ~\frac{u}{G ~\rho ~R} ~,
\nonumber \\
\beta_{E} &=& \frac{3}{16~ \pi} ~\frac{L_{\star} ~\overline{Q} ~'_{pr}}{G ~M_{\star} ~c ~\rho ~R} ~.
\end{eqnarray}
The quantity $\beta_{ESWR}$ is the relevant quantity which has to be used as the $\beta$ parameter in the initial values
of orbital elements presented by Kla\v{c}ka (2004 - Eqs. 58-59, or, as a special case, Eqs. 60-63).

Formally, inserting  Eq. (\ref{stars-3}) into Eq. (\ref{celmech8-star-ratio-e0-2}),
\begin{eqnarray}\label{celmech8-star-ratio-e0-3c}
\frac{\tau_{SW}}{\tau_{E}} &=& \frac{\overline{Q} ~'_{pr}}{Q_{wind}} ~\frac{L_{\star}}{\dot{M}_{\star} ~c^{2}}  ~.
\end{eqnarray}

Eq. (\ref{celmech8-star-ratio-e0-2}), the second of Eqs. (\ref{celmech8-star-ratio-e0-3}) and Eq. (\ref{stars-3}) yield
\begin{eqnarray}\label{force-ration-Q-wind}
\frac{\overrightarrow{F}_{SW} \cdot \overrightarrow{e}_{R}}{\overrightarrow{F}_{P-R} \cdot \overrightarrow{e}_{R}}
\frac{c}{u} &\doteq& \frac{\dot{M}_{\odot} ~c^{2}}{L_{\odot}} ~\frac{Q_{wind}}{\overline{Q} ~'_{pr}} ~
\frac{\dot{M}_{\star}}{\dot{M}_{\odot}} \left ( \frac{L_{\star}}{L_{\odot}} \right )^{-1} ~.
\end{eqnarray}

The first of Eqs. (\ref{beta-eta-kappa-star}), the second of Eqs. (\ref{celmech8-star-complete}) and Eq. (\ref{stars-3})
yield
\begin{eqnarray}\label{celmech8-star-complete-sw}
\tau_{SW} ( e_{\beta in} \ll 1 ; e_{\beta} ; \eta_{1} ; \eta_{2} ) &=& \frac{1}{{Q}_{wind}}
\frac{m}{\dot{M}_{\star} R^{2}} \left ( a_{\beta in} \right )^{2} ~.
\end{eqnarray}

\subsubsection{Numerical calculations}
Eq. (\ref{celmech8-star-ratio-e0-2}) yields
\begin{eqnarray}\label{celmech8-star-ratio-e0-2-num}
\frac{\tau_{SW}}{\tau_{E}} &=& \frac{\overline{Q} ~'_{pr}}{\eta_{2}}
\doteq \frac{5}{7} ~\overline{Q} ~'_{pr}
\end{eqnarray}
for the Solar System. This corresponds to 20\% of the value presented by Plavchan et al. (2005 - text below Eq. 3 on p. 1166).

Let us calculate the numerical value of $Q_{wind}$ given by Eq. (\ref{stars-3}).
We can use $L_{\odot}$ $=$ 3.824 $\times$ 10$^{26}$ $W$ and
$\dot{M}_{\odot}$ $\doteq$ 2.5 $\times$ 10$^{-14}$ $\mbox{M}_{\odot} ~\mbox{yr}^{-1}$
(see, e.g., $http://en.wikipedia.org/wiki/Solar\_{}wind$).
If we would use $\eta_{2}$ $=$ 0.38 (Kla\v{c}ka et al. 2012), then we would obtain $Q_{wind}$ $\doteq$ 1.0, which is consistent
with the conventionally used value (see, e.g., Strubbe and Chiang 2006 - p. 654). However, if we use the physical value
$\eta_{2}$ $=$ 1.4, then $Q_{wind}$ $\doteq$ 3.8. The values
$\dot{M}_{\odot}$ $\doteq$ 2.0 $\times$ 10$^{-14}$ $\mbox{M}_{\odot} ~\mbox{yr}^{-1}$
(e.g., Plavchan et al. 2009 -- Sec. 4.4.3) and $\eta_{2}$ $=$ 1.4 lead to
\begin{eqnarray}\label{stars-4}
Q_{wind} &\doteq& 4.7 ~.
\end{eqnarray}

If we would use the approximation $\tilde{\eta}_{\odot 2} / \tilde{\eta}_{\star 2}$ $\doteq$ 1
in Eq. (\ref{celmech8-star-ratio-e0-4}), then we would obtain
\begin{eqnarray}\label{celmech8-star-ratio-e0-4-new}
\frac{\tau_{SW}}{\tau_{E}} &\doteq& \frac{5}{7} ~\overline{Q} ~'_{pr} ~\frac{L_{\star}}{L_{\odot}}
~\frac{\dot{M}_{\odot}}{\dot{M}_{\star}} ~,
\end{eqnarray}
which is different from the conventional approach using the value 3 instead of 5/7
(see, e.g., Plavchan et. al 2009 - Eq. A10).

Eq. (\ref{force-ration-Q-wind}) leads to
\begin{eqnarray}\label{force-ration-Q-wind-num}
\frac{\overrightarrow{F}_{SW} \cdot \overrightarrow{e}_{R}}{\overrightarrow{F}_{P-R} \cdot \overrightarrow{e}_{R}}
\frac{c}{u} &\doteq& \frac{\eta_{2}}{\overline{Q} ~'_{pr}} ~
\frac{\dot{M}_{\star}}{\dot{M}_{\odot}} \left ( \frac{L_{\star}}{L_{\odot}} \right )^{-1} ~,
\nonumber \\
&\doteq& 0.3 ~\frac{Q_{wind}}{\overline{Q} ~'_{pr}} ~
\frac{\dot{M}_{\star}}{\dot{M}_{\odot}} \left ( \frac{L_{\star}}{L_{\odot}} \right )^{-1} ~,
\nonumber \\
\eta_{2} &=& 1.4 ~,
\end{eqnarray}
if the values $L_{\odot}$ $=$ 3.824 $\times$ 10$^{26}$ $W$ and $\dot{M}_{\odot}$ $\doteq$ 2.0 $\times$ 10$^{-14}$ $\mbox{M}_{\odot} ~\mbox{yr}^{-1}$
are used. The second of Eqs. (\ref{celmech8-star-ratio-e0-3}) and Eq. (\ref{force-ration-Q-wind-num}) give
\begin{eqnarray}\label{force-ration-Q-wind-num-1}
\frac{\tau_{SW}}{\tau_{E}} &\doteq& \frac{10}{3} ~\frac{\overline{Q} ~'_{pr}}{Q_{wind}} ~ \frac{L_{\star}}{L_{\odot}}
\left (  \frac{\dot{M}_{\star}}{\dot{M}_{\odot}} \right )^{-1} ~.
\end{eqnarray}
We stress that Eq. (\ref{stars-4}) holds.

On the basis of Eqs. (\ref{celmech8-star-complete}) we can write
\begin{eqnarray}\label{celmech8-star-complete-z-cloud}
\frac{\tau ( e_{\beta in} ; e_{\beta} ; \eta_{1} ; \eta_{2} )}{\tau_{E} ( e_{\beta in} ; e_{\beta} )}
&\le& \frac{\tau ( e_{\beta in} \ll 1 ; e_{\beta} ; \eta_{1} ; \eta_{2} )}{\tau_{E} ( e_{\beta in} \ll 1 ; e_{\beta} )}
\nonumber \\
&\doteq& \frac{1}{1 ~+~ \eta_{2} / \overline{Q} ~'_{pr}} ~.
\end{eqnarray}
This formula can be applied to the zodiacal cloud in the Solar System. Using the approach presented by Fixsen and Dwek (2002),
Eq. (\ref{celmech8-star-complete-z-cloud}) yields
\begin{eqnarray}\label{celmech8-star-complete-z-cloud-num}
\frac{\tau ( e_{\beta in} ; e_{\beta} ; \eta_{1} ; \eta_{2} )}{\tau_{E} ( e_{\beta in} ; e_{\beta} )}
&\le&  \frac{1}{1 ~+~ \eta_{2} / \overline{Q} ~'_{pr}} \doteq \frac{1}{4} ~,
\end{eqnarray}
since $\eta_{2}$ $=$ 1.4 and $\overline{Q} ~'_{pr}$ $=$ 1/2 (Fixsen and Dwek 2002, p. 1014).
The mass-loss rate of the micron-sized dust particles in the zodiacal cloud is about
\begin{eqnarray}\label{zod-cloud}
\dot{M}_{d} = 24 \times 10^{10} ~\mbox{kg} ~\mbox{yr}^{-1} ~,
\end{eqnarray}
i.e. 4-times higher than the value found by Fixsen and Dwek (2002).
For a particle mass density of $\varrho$ $=$ 3 $\mbox{g} ~\mbox{cm}^{-3}$, a grain radius of 30 $\mu \mbox{m}$,
and a radiation pressure efficiency $\overline{Q} ~'_{pr}$ $=$ 0.5, we find that the
lifetime for a particle at 1 AU is about 3.3 $\times$ 10$^{4}$ yr instead of 10$^{5}$ yr presented by Fixsen and Dwek (2002).

We now turn to an application to M dwarfs. On the basis of  Eqs. (\ref{stars-1-new}) and the data given
by Plavchan et al. (2005, Sec. 5. 6), we can write
\begin{eqnarray}\label{stars-1-new-num}
\beta_{ESWR} &\doteq&  \left ( x_{\beta E} ~\frac{g_{L}}{g_{M}} + x_{\beta wind} ~ \frac{f_{M}}{g_{M}} \right ) \frac{1}{R [\mu \mbox{m}]}
\frac{2 500}{\rho [\mbox{kg} ~\mbox{m}^{-3}]} ~,
\nonumber \\
x_{\beta E} &=& 2.30 \times 10^{-1} ~\overline{Q} ~'_{pr} ~,
\nonumber \\
x_{\beta wind} &=& 4.80 \times 10^{-2} ~\frac{Q_{wind}}{4.7} ~\frac{u \left [ \mbox{km} ~\mbox{s}^{-1} \right ]}{450} ~,
\nonumber \\
g_{L} &\equiv& \frac{L_{\star}}{0.1~L_{\odot}} ~,
\nonumber \\
g_{M} &\equiv& \frac{M_{\star}}{0.1~M_{\odot}} ~,
\nonumber \\
f_{M} &\equiv& \frac{\dot{M}_{\star}}{10~\dot{M}_{\odot}} ~.
\end{eqnarray}
Formula represented by Eqs. (\ref{stars-1-new-num}) offers higher values of $\beta_{ESWR}$ than the values discussed by Plavchan et al. (2005).
In comparison with the calculations of the authors, larger particles escape from the stars.
Moreover, one has to bear in mind that the statement below Eq. (\ref{stars-1-new}) holds: e.g., particle of
$\beta_{ESWR}$ $=$ ( 1 $-$ $e_{P}$) / 2 moves in parabolic orbit if ejected (with zero ejection speed)
at periastron of the parent body moving in an osculating ellipse with eccentricity $e_{P}$.
As an example, corresponding to the data discussed by Plavchan et al. (2005), we take the following physical values:
$\rho$ $=$ 2500 $\mbox{kg} ~\mbox{m}^{-3}$, $u$ $=$ 450 $\mbox{km} ~\mbox{s}^{-1}$, $\overline{Q} ~'_{pr}$ $=$ 1,
$Q_{wind}$ $=$ 4.7, $L_{\star}$ $=$ 0.1 $L_{\odot}$, $M_{\star}$ $=$ 0.5 $M_{\odot}$, $\dot{M}_{\star}$ $=$ $\dot{M}_{\odot}$.
Then, $R$ $=$ 0.04696 $\mu \mbox{m}$ / $\beta_{ESWR}$, which yields
$R$ $=$ 4.70 $\times$ 10$^{-2}$ $\mu \mbox{m}$ for $\beta_{ESWR}$ $=$ 1, and, $R$ $=$ 2.35 $\times$ 10$^{-1}$ $\mu \mbox{m}$ for $e_{P}$ $=$ 0.6.
These values are 235, and, 1175-times greater than the value 2 $\times$ 10$^{-4}$ $\mu \mbox{m}$ presented by Plavchan et al. (2005)
as a blowout radius.

A relation between the luminosity of the infrared excess due to dust $L_{1R}$ $\approx$ $(\nu L_{\nu})_{IR}$ and the rate
at which mass is being removed from parent bodies and converted into dust, $\dot{M}_{d}$, is
\begin{eqnarray}\label{dot-M-d-1}
\dot{M}_{d} &=& \frac{C_{0} ~L_{1R}}{c^{2}} ~ \left ( 1 ~+~ \frac{Q_{wind}}{\overline{Q} ~'_{pr}} ~
\frac{\dot{M}_{\star} ~c^{2}}{L_{\star}} \right ) ~,
\end{eqnarray}
where $C_{0}$ is a numerical constant of order unity and it depends on the assumed initial dust distribution
relative to the inner radius at which the dust sublimates (see, e.g., Plavchan et al. 2005, where also
the value $C_{0}$ $=$ 4 is presented). Eq. (\ref{dot-M-d-1}) is consistent with Kla\v{c}ka (2013), but while
$Q_{wind}$ $\doteq$ 9/4 for the Maxwell-Boltzmann distribution, the value $Q_{wind}$ $\doteq$ 4.7 holds
for the $\kappa-$distribution - compare Eq. (\ref{stars-4}).

\section{Conclusion}
The physics of the solar/stellar drag is given by Eqs. (\ref{sw-force-total-fin-kappa2})-(\ref{sw-force-total-fin-kappa2-zeta}).
The solar/stellar wind corpuscles strike the dust particle orbiting the Sun/star and they act as a drag force. The drag coefficient
$c_{D}^{tot}$ depends on the velocity distribution holding for the wind corpuscles.

Equation of motion of a homogeneous spherical body under the action of the gravity of the Sun and
the solar electromagnetic and corpuscular radiation is given by
Eqs. (\ref{sw-force-total-fin-kappa2})-(\ref{sw-force-total-fin-kappa2-zeta}), where the data from Sec. 3 have to be used.
As an approximation,
Eqs. (\ref{accel-total-1st-order1}), (\ref{auxiliary1}) and (\ref{auxiliary2})
can be used, if also the P-R effect is considered.
On the basis of Eq. (\ref{accel-total-1st-order2})
we can write the most simple form of the equation of motion (to terms of order $v/c$ and $v/u$):
\begin{eqnarray}\label{accel-total-1st-order2-con}
\frac{d \overrightarrow{v}}{dt} &\doteq& \beta ~\frac{G~M_{\odot}}{r^{2}}
\nonumber \\
& & \times \left [ 1 + \frac{\eta_{2}}{\overline{Q} ~'_{pr}} ~ \frac{u}{c}
~-~ \left ( 1 +  \frac{\eta_{1}}{\overline{Q} ~'_{pr}} \right )
\frac{\overrightarrow{v} \cdot \overrightarrow{e}_{R}}{c}  \right ] \overrightarrow{e}_{R}
\nonumber \\
& & -~ \beta ~\frac{G~M_{\odot}}{r^{2}} \left ( 1 ~+~ \frac{\eta_{2}}{\overline{Q} ~'_{pr}} \right ) \frac{\overrightarrow{v}}{c}
\nonumber \\
& & -~ \frac{G~M_{\odot}}{r^{2}} ~\overrightarrow{e}_{R} ~,
\end{eqnarray}
if also gravity of the Sun is used.
The obtained results are based on the observational fact that $\kappa$-distribution holds for the solar
wind corpuscles. The term $( \eta_{2} / \overline{Q} ~'_{pr} ) u/ c$ cannot be neglected for the time-variable solar wind.

The conventional approach uses $\eta_{1}$ $=$ $\eta_{2}$ $\in$ $\langle 0.2, 0.3 \rangle$, the Maxwell-Boltzmann velocity distribution
of solar wind corpuscles leads to $\eta_{1}$ $=$ $\eta_{2}$ $=$ 2/3, because the relevant contributions of the solar wind action
contain also the sputtering and reflection components in addition to direct impact. The physical result based on the observed
$\kappa-$distribution yields $\eta_{1}$ $\doteq$ 1.1, $\eta_{2}$ $\doteq$ 1.4 for the Solar System.
The relation $\beta$ $=$ 5.760 $\times$ 10$^{2}$ $(L_{\star} / L_{\odot} )$ $( M_{\odot} / M_{\star} )$
$\overline{Q} ~'_{pr} / ( R [\mu m] \rho [kg~m^{-3}] )$ holds for the homogeneous particle of radius $R$ and mass density $\rho$. Moreover,
the following approximations can be used: $\eta_{i \star}$ $\doteq$ $\eta_{i \odot}$ $(u_{w \star} / u_{w \odot})$ $[n_{p \star} (r_{0} )/ n_{p \odot}  (r_{0} )]$
$(L_{\odot} / L_{\star})$, $i$ $=$ 1 and 2, where $u_{w}$ is the wind speed, $n_{p} (r_{0} )$ is the concentration of protons at a given distance $r_{0}$
and $L$ is the luminosity of the central object. As for the exact approach, see Eqs. (\ref{tilde-eta-1-2-3}), (\ref{beta-eta-kappa}) and (\ref{beta-eta-kappa-star}).

As for the secular evolution, the most simple case is described by Eqs. (\ref{celmech5})-(\ref{celmech6}).
Quantitative result comparing the importance of the solar wind with respect to the P-R effect, as for the secular evolution,
is presented by Eq. (\ref{celmech8-star-ratio-e0-2}) corresponding to the approximation of near circular orbit.
The solar wind effect is 1.4-times more important than the P-R effect, if $\overline{Q} ~'_{pr}$ $=$ 1, as for
the secular orbital evolution of the spherical dust particle.

Owing to the transversal component of the P-R effect and the effect of the solar wind, the orbits
of Solar System grains collapse in
\begin{eqnarray}\label{celmech-stars-1}
\tau_{PR+sw} &\doteq& \frac{7.0 \times 10^{6} ~\mbox{yrs}}{\overline{Q} ~'_{pr} + \eta_{2}} ~R~ \rho ~r^{2}
\nonumber \\
&\doteq& \frac{400 ~\mbox{yrs}}{\beta \left ( 1 + \eta_{2} / \overline{Q} ~'_{pr} \right )} ~r^{2} ~,
\nonumber \\
\eta_{2} &\doteq& 1.4 ~,
\end{eqnarray}
where $R [\mbox{cm}]$ is particle's radius, $\rho [\mbox{g}~\mbox{cm}^{-3}]$ its mass density, $r [\mbox{AU}]$
is heliocentric distance of the particle (trajectory is approximated by a circle - near circular orbit). This holds under the assumption
of the radial solar wind.
For a particle mass density of $\varrho$ $=$ 3 $\mbox{g} ~\mbox{cm}^{-3}$, a grain radius of 30 $\mu \mbox{m}$,
and a radiation pressure efficiency $\overline{Q} ~'_{pr}$ $=$ 0.5, we find that the
lifetime for a particle at 1 AU is about 3.3 $\times$ 10$^{4}$ yr instead of 1.3 $\times$ 10$^{5}$ yr found by Fixsen and Dwek (2002).

The non-radial component of the solar wind velocity ($\gamma_{T}$ $\ne$ 0 in Eq. \ref{sw-force-total-fin-kappa2}), the change of the
solar wind properties during the solar cycle and the decrease of the particle's mass enhance the difference
between the solar wind action and the P-R effect.

The equation of motion represented by the above discussed forms, including the simple Eq. (\ref{accel-total-1st-order2-con}),
may significantly change our understanding of the long-term orbital evolution of dust particles, since the total radiation effect is
more important than it has been considered up to now. This holds both for the Solar System and surroundings of other stars with
stellar winds, see also Sec. 5.2 (e.g., the dimensionless cross section the dust grain presents to wind pressure is $Q_{wind}$ $\doteq$ 4.7,
instead of the conventionally used value 1.0). Maybe, initial stages of creation of planetary systems are more rapid.
As an application we can also mention that more abundant sources of dust grains, e.g., asteroids and short periodic comets, are required in the Solar System
(about 4-times of the conventional ideas). Other applications concern orbital evolution in mean-motion orbital resonances with planets,
possible capture of interstellar dust grains in the Solar System, etc. .

\section*{Acknowledgement} This work was supported by the Scientific Grant Agency VEGA No. 1/0670/13.

\end{document}